\documentclass[oldversion]{aa}

\usepackage{txfonts}
\usepackage{graphicx}
\usepackage{natbib}

\newcommand{\Halpha}{\ensuremath{{\mathrm H}\alpha}}
\newcommand{\Hbeta}{\ensuremath{{\mathrm H}\beta}}
\newcommand{\cycday}{\ensuremath{\mathrm{d}^{-1}}}
\newcommand{\kms}{\,\mbox{$\mathrm{km}~\mathrm{s}^{-1}$}}
\newcommand{\Teff}{\ensuremath{T_\mathrm{eff}}}
\newcommand{\Kelvin}{\ensuremath{\,\mathrm{K}}}
\newcommand{\hei}{\ion{He}{i}}
\newcommand{\siii}{\ion{Si}{ii}}
\newcommand{\RV}{{\sl RV}}
\newcommand{\hve}{{HR\,1847}}
\newcommand{\hvea}{{HR\,1847A}}
\newcommand{\hveb}{{HR\,1847B}}
\newcommand{\compa}{component~A}
\newcommand{\compb}{component~B}
\newcommand{\vsinia}{\ensuremath{(\varv \sin i)_\mathrm{A}}}
\newcommand{\vsinib}{\ensuremath{(\varv \sin i)_\mathrm{B}}}
\newcommand{\Hipparcos}{{\rm HIPPARCOS}}
\newcommand{\HIPPARCOS}{\Hipparcos}
\newcommand{\HEROS}{{\em HEROS}}
\newcommand{\MIDAS}{{\tt MIDAS}}
\newcommand{\IRAF}{{\tt IRAF}}
\newcommand{\SPEFO}{{\tt SPEFO}}
\newcommand{\FOTEL}{{\tt FOTEL}}

\newcommand{\PERIOD}{{\tt PERIOD}}
\newcommand{\ROTIN}{{\tt ROTIN3}}

\newcommand{\lastA}{25 February 2008}
\newcommand{\lastB}{22 March 2008}
\newcommand{\lastRA}{30 August 2007}
\newcommand{\lastRB}{30 August 2007}
\newcommand{\numberA}{68}
\newcommand{\numberB}{38}

\begin{document}
\title{Spectroscopic analysis of the B/Be visual binary
\hve\thanks{Based on observations with the Ond\v{r}ejov and Rozhen
2-m telescopes.}}
\author{J.\,Kub\'{a}t\inst{1},
        S.\,M.\,Saad\inst{2},
	A.\,Kawka\inst{1},
	M.\,I.\,Nouh\inst{2},
	L.\,Iliev\inst{3},
	K.\,Uytterhoeven\inst{4},
	D.\,Kor\v{c}\'akov\'a\inst{1},
	P.\,Hadrava\inst{5},
	P.\,\v{S}koda\inst{1},
	V.\,Votruba\inst{1,6},
	M.\,Dov\v{c}iak\inst{5},
 	M.\,\v{S}lechta\inst{1}
	}

\offprints{J.~Kub\'at, \\ \email{kubat@sunstel.asu.cas.cz}}

\authorrunning{J.~Kub\'{a}t et al.}

\institute{Astronomick\'y \'ustav,
	Akademie v\v{e}d \v{C}esk\'e republiky,
	CZ-251 65 Ond\v{r}ejov, Czech Republic
	\and
	National Research Institute of Astronomy and Geophysics,
	11421 Helwan, Cairo, Egypt
	\and
	Institute of Astronomy, Bulgarian Academy of Sciences,
	72 Tsarigradsko Shossee Blvd., BG-1784 Sofia, Bulgaria
	\and
	Laboratoire AIM, CEA/DSM-CNRS-Universit\'e Paris Diderot, CEA,
	IRFU, SAp, Centre de Saclay, F-91191 Gif-sur-Yvette, France
	\and
	Astronomick\'y \'ustav,
	Akademie v\v{e}d \v{C}esk\'e republiky,
	Bo\v{c}n\'{\i} II 1401, CZ-141 31 Praha 4, Czech Republic
	\and
	\'Ustav teoretick\'e fyziky a astrofyziky P\v{r}F MU,
	Kotl\'a\v{r}sk\'a 2, CZ-611 37 Brno, Czech Republic
	}

\date{draft version \today}

\abstract{We studied both components of a slightly overlooked visual
binary {\hve} spectroscopically to determine its basic physical and
orbital parameters.
Basic stellar parameters were determined by comparing synthetic spectra
to the observed echelle spectra, which cover both the optical and
near-IR regions.
New observations of this system used the Ond\v{r}ejov and Rozhen 2-m
telescopes and their coud\'e spectrographs.
Radial velocities from individual spectra were measured and then
analysed with the code {\FOTEL} to determine orbital parameters.
The spectroscopic orbit of {\hvea} is presented for the first time.
It is a single-lined spectroscopic binary with a B-type primary, a
period of 719.79 days, and a highly eccentric orbit with $e=0.7$.
We confirmed that {\hveb} is a Be star.
Its {\Halpha} emission significantly decreased from 2003 to 2008.
Both components have a spectral type B7--8 and luminosity class IV--V.}


\maketitle

\section{Introduction}
{\hve} (HD\,36408, BD\,+16$^{\circ}$794, HIP\,25950, ADS\,4131)
is a bright visual binary consisting of two B type stars.
Throughout this paper, the component at
$\alpha(\mathrm{J}2000)=05\,32\,14.14$,
$\delta(\mathrm{J}2000)=+17\,03\,29.3$ will be denoted as
\object{{\hvea}} (or \compa) and the one at
$\alpha(\mathrm{J}2000)=05\,32\,14.56$,
$\delta(\mathrm{J}2000)=+17\,03\,21.8$ as \object{{\hveb}} (or \compb).
The first observations of this binary, which were obtained on
31 December 1782, were reported by William Herschel \citep[see][star
III.93]{catbin2}.
He observed this binary again on 21 January 1800 \citep[see][star
124]{newbin145}.

Athough this visual binary consists of two bright stars and
\citet{DAOnote} recommended further observations, it was not studied
much during the last century.
Since {\hve} is both an X-ray source \citep*{ROSAT} and an IRAS source
\citep{ir101}, and considering its brightness, lack of observations, and
sometimes confusing catalogue information, it has become quite an
interesting object for more detailed study.

\section{Summary of known properties of {\hve}}

In this section we summarize the information currently available for
both components of {\hve}, which is scattered throughout the
astronomical literature, online catalogues, and databases.

\subsection{Multiplicity}

\subsubsection{Angular separation}\label{relpos}

Angular separation $\rho$ and relative position $\theta$ of the
components of {\hve} have already been measured by Friedrich Georg
Wilhelm \citet[][star No.\,730]{mensurae} as an average of 4
measurements obtained in 1829 and 1832 ($\rho=9\farcs81,
\theta=141\fdg82$).
Later measurements resulted in $\rho=9\farcs53, \theta=140\fdg2$
\citep{acdouble}, $\rho=9\farcs7, \theta=140\fdg9$ \citep{dynpar},
$\rho=9\farcs65, \theta=140\fdg85$ \citep[from \Hipparcos]{hipparcos},
$\rho=9\farcs65\pm0\farcs006, \theta=140\fdg5$ \citep[from the Tycho
catalogue]{tychodouble}, $\rho=9\farcs61, \theta=140\fdg8$ \citep[from
speckle interferometry]{speckle9}.
The angular separation, as reported by \citet{double2} and \citet{Bebin}
is $9\farcs6$.

\subsubsection{Radial velocity variations}

The first radial velocity measurements for both components were reported
by \citet{DAObin3} and they showed variability.
\cite{DAO100} suggested that both components are spectroscopic binaries.
\citet{HeRV} designated {\hvea} (No.\,126 in his Table\,II) as a
spectroscopic binary with an unknown orbit using only one spectrum
secured at the Dominion Astrophysical Observatory (DAO) in 1919 by Otto
Struve.
Component A was also reported to be a spectroscopic binary by
\citet*{double1} and \citet*{ALe02}.
However, no attempt has been made to determine the orbital parameters
until now.

\subsection{Distance}

The {\Hipparcos} parallax of {\hve} is $2.92\pm1.57$\,mas
\citep{hipparcos}, which corresponds to a distance of $342\pm184$pc, but
the large error reduces the reliability of this distance estimate.
However, the new edition of the {\Hipparcos} catalogue
\citep{hippnewred} gives a parallax value of $1.09\pm 2.04$\,mas, which
is unable to provide any information about the distance owing to the
very uncertain parallax.
Individual parallax measurements of the A and B components are not
available.

\subsection{Magnitudes and photometry}\label{magphot}

\begin{table}
\caption{{\Hipparcos} and $UBVRIJHK$ photometry of {\hve}.}
\label{magtab}
\begin{tabular}{|c|c|c|c|c|}
\hline
& {\hvea} & {\hveb} & source \\
\cline{1-3}
band & mag & mag &  \\
\hline
$H_p$ & $6.088\pm0.007$ & $6.455\pm0.010$ & \HIPPARCOS \\
$B_T$ & $6.036\pm0.008$ & $6.468\pm0.004$ & \HIPPARCOS \\
$V_T$ & $6.057\pm0.009$ & $6.458\pm0.004$ & \HIPPARCOS \\
\hline
$U$ & $5.823\pm0.030$ & $6.337\pm0.030$ & EXPORT \\
$B$ & $6.103\pm0.037$ & $6.527\pm0.037$ & EXPORT \\
    & $6.070\pm0.020$ & $6.520\pm0.020$ & Maidanak \\
$V$ & $6.090\pm0.100$ & $6.490\pm0.077$ & EXPORT \\
    & $6.090\pm0.020$ & $6.510\pm0.020$ & Maidanak \\
$R$ & $6.037\pm0.100$ & $6.410\pm0.077$ & EXPORT \\
    & $6.050\pm0.020$ & $6.460\pm0.020$ & Maidanak \\
$I$ & $6.050\pm0.150$ & $6.437\pm0.133$ & EXPORT \\
$J$ & $5.964\pm0.019$ & $6.280\pm0.024$ & 2MASS \\
$H$ & $6.022\pm0.021$ & $6.275\pm0.021$ & 2MASS \\
$K$ & $6.016\pm0.027$ & $6.282\pm0.016$ & 2MASS \\
\hline
\end{tabular}

\vspace{1mm}
{\HIPPARCOS} -- \cite{hipparcos},
EXPORT -- \cite{Oud01},
Maidanak -- \cite{vis82},
2MASS -- \cite{2masspubl}.
\end{table}

The first reliable magnitude values ($V_A=6.07$, $V_B=6.44$) were
published by \cite{acdouble}.
She also gives the combined AB magnitude as $5.49$.
Maybe that only the combined magnitude value was given in her earlier
spectral classification \citep{acclass} and in the HD catalogue
\citep{HD456} caused its being incorrectly but often quoted as the
magnitude of the A component until now.
A similar value is also listed in the Bright Star Catalogue
\citep[$V=5.46$,][]{bsc5}.

The photometry of {\hve} was measured by several authors always during
observing runs devoted to large sample of stars.
\cite{vejce} measured $uvby\beta$ photometry of early type stars and
derived values $V_A=6.1$ and $V_B=6.5$.
\cite{double2} measured $uvby\beta$ photometry of visual binaries
including {\hve}.
He derived corresponding $V$ magnitudes of both components from
the $y$ filter values as $V_A=6.091\pm0.019$ and $V_B=6.510\pm0.023$.
$BVR$ observations of visual binaries at Mount Maidanak were performed
by \cite{vis82}, and he obtained $V_A=6.09\pm0.02$ and
$V_B=6.51\pm0.02$.
The EXPORT $UBVRI$ photometry \citep[observed by the Nordic Optical
Telescope,][]{Oud01} gave values $V_A=6.09\pm0.10$ and
$V_B=6.49\pm0.07$.
Recent determinations of magnitudes were derived from {\Hipparcos}
photometry \citep[see][]{hipparcos}.
With the {\Hipparcos} $V_T$ and $B_T$ magnitudes, \cite{miliony} derived
the magnitudes $V_A=6.056\pm0.008$ and $V_B=6.451\pm0.004$, which do not
differ too much from the first values of Cannon.
A more complete list of photometric magnitudes of both components of
{\hve} is given in Table~\ref{magtab}.

\subsection{Polarization}

\citet{Oud01} found the value of $V$ band polarization $P_V =
0.0063\pm0.0004$ for the A component, and $P_V = 0.0063\pm0.0002$ for
the B component.

\subsection{Spectral types and luminosity classes}

The early classification of both components was \emph{`egregie alba'}
(very white) by \cite{mensurae}.
Later, \citep{acdouble} classified both stars as B9.

Using his own observations in the period from April 1954 to March 1955
at the Yerkes Observatory, \cite{b8a2} determined the spectral type of
{\compa} as B7IV and of {\compb} as B8IV.
From his own observations with the Perkins telescope, \cite{rotdbl}
determined spectral types B7III for {\compa} and B7IV for {\compb}.
\cite{class} determined spectral types B8III and B8V for components A
and B, respectively, using spectra from MacDonald and Yerkes
Observatories.
\cite{vis} observed both components at Cerro Tololo Inter-American
Observatory and obtained spectral type B7III for component A and B8IV
for B.
Finally, \cite{EXPORT} changed the Bright Star Catalogue classification
of {\hvea} from B7IIIe to B5V.

\subsubsection{Emission}

\cite{halphot} found that {\compb} shows {\Halpha} emission using
{\Halpha} photometry, which was later spectroscopically confirmed by
\cite{emise}.
The emission sign, which appeared at the spectral type of {\compa} in
both the Bright Star Catalogue and the Simbad database, is incorrect, as
{\Halpha} emission was never reported for {\hvea}.

\subsection{Rotation}

Rotational velocities of both components were first determined by
\cite{rotdbl}.
The projected rotational velocity of {\compa} was measured as $\vsinia =
50\kms$, while {\compb} was found to be rotating rapidly, $\vsinib =
300\kms$.
\cite{vis} determined $\vsinia = 60\kms$ and $\vsinib = 300\kms$.
\cite{ALe02} refined the values of the rotational velocities to $\vsinia
= 45\kms$ and $\vsinib = 200\kms$ using coud\'e spectra obtained at the
Kitt Peak 0.9-m telescope.

\subsection{Variability}

Slightly different values of magnitude obtained at different times using
different instruments, which are listed in the Section \ref{magphot},
cannot be considered as firm proof of variability.
The only available homogeneous set of observations was obtained by the
{\Hipparcos} satellite.
Using {\Hipparcos} photometry, \citet*{Pe04} determined the
characteristic time scale of variability for {\hve} as 0.9 days.
They conclude that it is a classical Be star, but it is not clear which
component they are refering to, since both components A and B have
{\Hipparcos} photometry available.
In addition, data for both components also include several values that
correspond to the combined magnitude of both components.
Since it is not clear that they excluded the apparently wrong values
(those corresponding to combined A+B magnitude), their result is
questionable.
We performed an independent search for variability and could not confirm
the variability time scale found by \citet{Pe04}.

\subsection{X-rays}\label{xpaprsky}

{\hve} has also been identified as an X-ray source by ROSAT
\citep{ROSAT} with an X-ray luminosity $\log L_X=29.64$ (in the paper
version).
In the online version of the same catalogue (their Table~2), this value
is slightly different, $\log L_X=29.79$.
Both corresponding values of $\log (L_X/L_\mathrm{bol})=-6.82$ (or
$-6.79$ in the online version) correspond to a typical $\log
(L_X/L_\mathrm{bol})$ relation for B-type X-ray emitters
\citep[cf.][]{ROSATrel}.
It is, of course, a question of which star of the pair is the X-ray
source, if not both.
The position of the ROSAT source 1RXS~J053214.9+170319 suggests it is
more probably coincident with the star TYC~1301-1942-1={\hveb}
\citep{Xcatalog}; however, the positional error is $18^{\prime\prime}$
\citep{ROSATfaint}, so {\hvea} also lies within the error circle.

\subsection{IR excess}\label{infrak}

In their study of IR excess of 101 Be stars \citet{ir101} found IRAS
magnitudes $[12]=4.95$, $[25]=2.22$, $[60]=-0.92$, which places {\hve}
in the $[12]-[25]/[12]-[60]$ colour-colour diagram far away from the
region where almost all Be stars are located.
\citeauthor{ir101} suggest there is a reflection nebula close to {\hve}.
\citet{redbe} associate {\hve} with the IRAS SSSC source X0501+589.
\citet{IRAS} suggest that {\hve} is a good candidate for a Herbig Ae/Be
star with a very large or a very cool circumstellar disk, but
\citet{YSO} did not find any CO emission for this star.

Again, it is not clear which star has IR excess, but most likely it is
\compb, which was found to be a Be star.
There is also a small possibility that both stars are Be stars, with
{\compa} not in emission at that moment.
Further observations may shed a light on it.
\citet{ir101} attributed the spectral type B7IIIe to {\hve}, which
corresponds to \compb, and $\varv\sin i=55\kms$, which corresponds to
\compa.

\subsection{Cluster membership}

Although {\hve} is located in the same region as the open cluster
Collinder~65 (A component is \#771, B component \#772, according to the
WEBDA\footnote{The Galactic and Magellanic Clouds open cluster database
WEBDA is available at http://obswww.unige.ch/webda/} database), its
membership is improbable mainly because to the different proper motion
\citep{Coll65}.

\section{Observations and data reduction}

The data available for this study consist of several data sets of
electronic spectra mostly centred on the {\Halpha} region:

\paragraph{Component A}

\begin{itemize}

\item
{\numberA} spectra in the spectral range 6250$-$6770\,{\AA} obtained
with a CCD SITe ST-005 800$\times$2000\,pix camera attached to the
coud\'{e} spectrograph of the 2m-telescope in Ond\v{r}ejov (Czech
Republic).
The spectra were obtained between 18 October 2003 and {\lastA}.
Spectra were reduced using {\IRAF}\footnote{IRAF is distributed by the
National Optical Astronomy Observatories, which are operated by the
Association of Universities for Research in Astronomy, Inc., under
cooperative agreement with the National Science Foundation.}.

\item
One spectrum obtained on 22 March 2003 with the red (5850$-$8450\,\AA)
and blue (3800$-$5650\,\AA) channels of the fiber-fed echelle
spectrograph {\HEROS} \citep[resolving power $\sim$ 20\,000, for its
brief description see][]{HERpopis}.
The spectrograph was attached to the Cassegrain focus of the
2m-telescope at the Ond\v{r}ejov Observatory.
All the basic data reduction processing was done using the {\HEROS}
pipeline written by Otmar Stahl and Andreas Kaufer as an extension of
the basic {\MIDAS} echelle context (\citealp[see][]{hermidas},
\citealp[also][]{herred}).

\item
14 spectra obtained in the spectral range (6510$-$6608\,\AA) at Rozhen
Observatory (Bulgaria) using the coud\'e-spectrograph of the 2m RCC
telescope.
The CCD camera Photometrics AT200 with an SITe SI003AB 1024x1024 CCD
chip (24x24 $\mu$m pixel size) was used in the f/9.5 camera of the
spectrograph to provide spectra with a spectral resolution of 36000 in
the H$\alpha$ region.
Spectra were obtained between 24 October 2004 and {\lastRA} and reduced
with the {\MIDAS} package.

\item
One spectrum in the spectral range 6290$-$6745\,{\AA} obtained at
Observat\'orio do Pico dos Dias (OPD, Brasil) and published by
\cite{brazilie}.
For details and information about the reduction process, see the
reference above.

\end{itemize}

In addition to the electronic spectra, we also used the radial velocity
measurements of \cite{DAObin3}.
For comparison purposes, several values of published radial velocities
were used from the works of \citet{HeRV} and \cite{rvfrance}.

\paragraph{Component B} spectra came from the same instruments as
mentioned in the preceding part describing spectra of {\compa}.

\begin{itemize}

\item
{\numberB} spectra obtained using the Ond\v{r}ejov coud\'e spectrograph
between 17 October 2003 and {\lastB}.

\item One {\HEROS} spectrum obtained at the Ond\v{r}ejov Observatory,
one night after the {\hvea} spectrum, on 23 March 2003.

\item
Three spectra between 28 August 2007 and {\lastRB} obtained at the
Rozhen Observatory.

\end{itemize}

\subsection{Radial velocity measurements}
\label{rvmereni}

Radial velocities were measured using the code {\SPEFO}, which was
developed by the late Dr. Ji\v{r}\'{\i} Horn \citep[see also][]{S96}.
The {\tt FITS} files obtained from {\IRAF} and {\MIDAS} were transformed
to the {\SPEFO} format, and then the radial velocities ({\RV}s) were
obtained interactively by means of the best match of the line profile
with its mirror.
The {\RV}s were obtained for hydrogen {\Halpha}, {\ion{He}{i}}
6678\,{\AA}, {\ion{Si}{ii}} 6347\,{\AA} and 6371\,{\AA} lines.
{\RV} data for individual components are described later in this paper
in Sections \ref{secRVA} and \ref{secRVB}.

\section{Component A}

\subsection{The effective temperature and gravity}\label{tempgrA}

The spectral region observed by {\HEROS} covered the range
3800$-$8620\,{\AA} (see Figs.\,\ref{HERAblue} and \ref{HERAred}) and was
used to determine the spectroscopic effective temperature and surface
gravity.
The line spectrum of the star is quite poor.
Balmer lines up to H10 and some infrared lines are seen.
No emission is present in the Balmer lines and in the lines of other
metal ions.

The effective temperature and surface gravity were determined by
comparing the observed spectra to model spectra.
We used the ATLAS9 LTE line blanketed model atmospheres, which were
calculated by \citet{Kur13} assuming solar metallicity and
microturbulent velocity 2{\kms}.
All synthetic spectra were convolved (using the code {\ROTIN} by
I.~Hubeny) with the Gaussian function having FWHM=0.25 to reduce the
resolution of the synthetic spectra to the observed resolution.
We used a $\chi^2$-fitting routine to compare the whole observed spectra
with the synthetic ones, and we obtained a best fit of
$\Teff=(12500\pm500)\Kelvin$ and $\log g=3.5\pm0.5$.
Following the temperature scale of \cite{classo8f6}, these values
correspond to spectral type B7 -- B8, and from the value of the surface
gravity, we estimate the luminosity class as IV -- V.
The quoted error bars correspond to the adopted steps of our model grid,
which are $500\Kelvin$ in temperature and 0.5 in $\log g$.

Using the code {\ROTIN}, we calculated a grid of rotationally broadened
spectra for $5\kms\le \varv\sin i \le 500 \kms$ with a step of 5\kms.
Then we applied the $\chi^2$ fitting routine again to determine the best
rotational velocity for this star.
To do this, we used the spectral line \ion{Mg}{ii} 4481\,{\AA}
\cite[recommended by][as a line free of pressure broadening]{grej} in
the fitting process.
We determined the projected rotational velocity of {\hvea} as $\varv
\sin i = (40\pm3)\kms$, which differs by 5{\kms} from the value of
\cite{ALe02}.
The error of our value of the rotational velocity was calculated
employing a 1$-\sigma$ error algorithm of \cite{htcas}.

\subsection{Radial velocities}
\label{secRVA}

\begin{figure}[t]
\resizebox{\hsize}{!}{\includegraphics{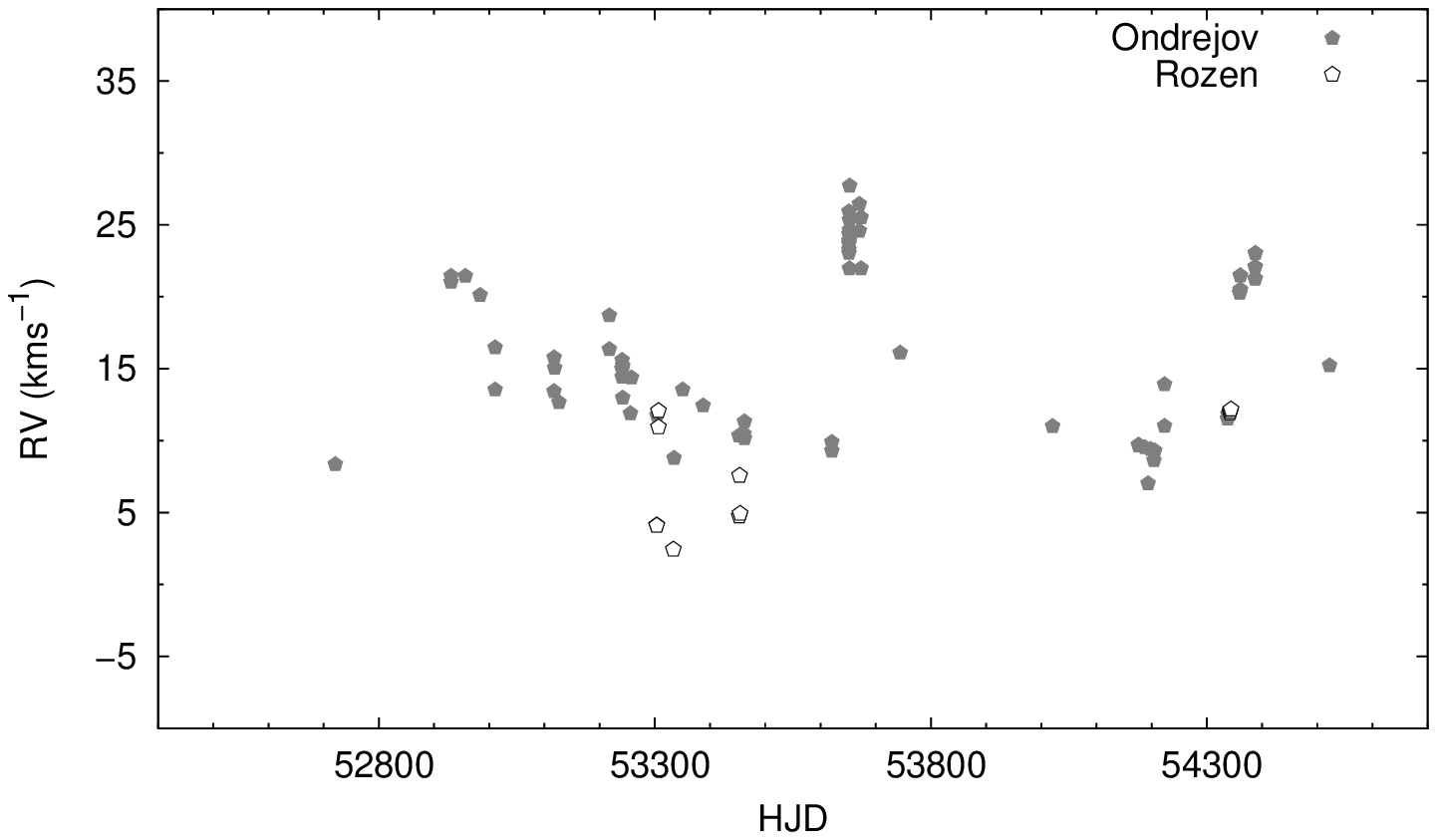}}
\resizebox{\hsize}{!}{\includegraphics{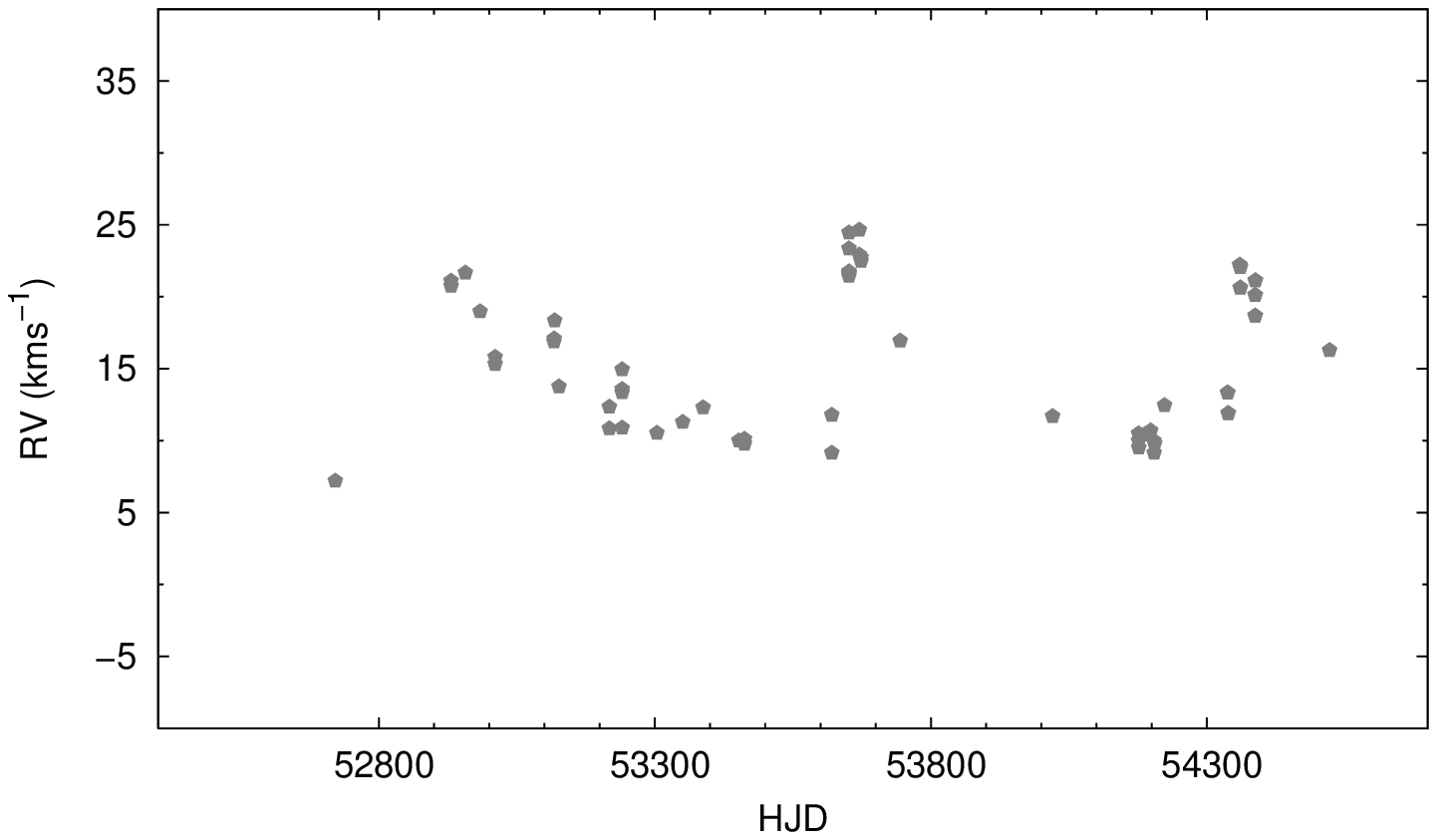}}
\caption {The radial velocity variations for hydrogen {\Halpha} (upper
panel) and {\hei} 6678\,{\AA} (lower panel) lines measured from the
Ond\v{r}ejov and Rozhen spectrograms as a function of HJD.
The {\hei} 6678\,{\AA} line is only present in the Ond\v{r}ejov spectra.
Full symbols represent Ond\v{r}ejov data, and open symbols stand for
Rozhen data.
All values are from Tables \ref{rvaond} and \ref{rvaroz}.}
\label{measrv}
\end{figure}

Tables\,\ref{rvaond} and \ref{rvaroz} display the line radial velocity
measurements that were derived using the methods described in Section
\ref{rvmereni}.
Columns 1, 2, and 3 represent the file identification, HJD, and the
heliocentric {\RV}s, respectively.
Column 4 lists the {\RV} measurements for the {\Halpha} line.
Columns 5, 6, and 7 list the {\RV}s measurements for {\hei} 6678\,{\AA},
{\siii} 6347\,{\AA}, and {\siii} 6371\,{\AA} (which are weak) absorption
lines, respectively.
The measured {\RV}s were shifted to the zero-point using a set of sharp
telluric absorption lines by means of the technique described in
\citet{Ho96}.

The {\RV}s of {\hvea} vary between 8{\kms} to 27{\kms} with a mean
velocity around 13{\kms}.
Figure~\ref{measrv} displays the measured {\RV}s for hydrogen {\Halpha}
and {\hei} 6678\,{\AA} as a function of time.
The plots indicate that our {\RV} measurement started with one value
just before the {\RV} maximum, then after the maximum, a clear decrease
in the {\RV} values is recorded until the {\RV} values reach their
minimum around HJD 2453620 with {\RV} values of 9$-$10\,{\kms}.
Then they quickly reached the next maximum value of 27\,{\kms} around
HJD~2453670.
Although only two cycles are covered by our observations, fortunately we
successfully recorded two epochs of rapid rise to the maximum, which
happened during a very short time (nearly 35 days) of the long cycle.
Rozhen observations confirmed the present distribution of {\RV}s.
Such a relatively short event could easily be missed. 

The {\RV} measurements that were obtained with different lines in the
{\Halpha} region show the same distribution with time, and only a small
difference in the maximum height can be recognized, as illustrated for
the \ion{He}{i} 6678\,{\AA} line (lower panel of Fig.\,\ref{measrv}).

\subsection{Period determination}

\begin{figure}
\resizebox{\hsize}{!}{\includegraphics{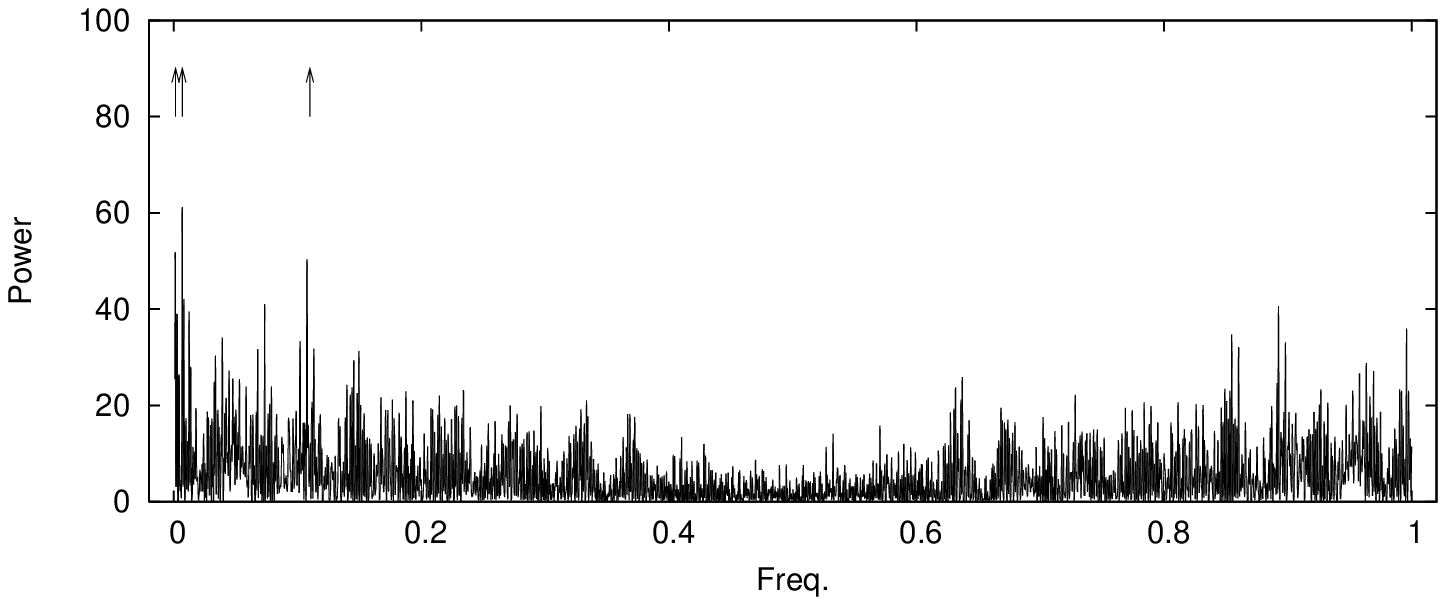}}
\resizebox{\hsize}{!}{\includegraphics{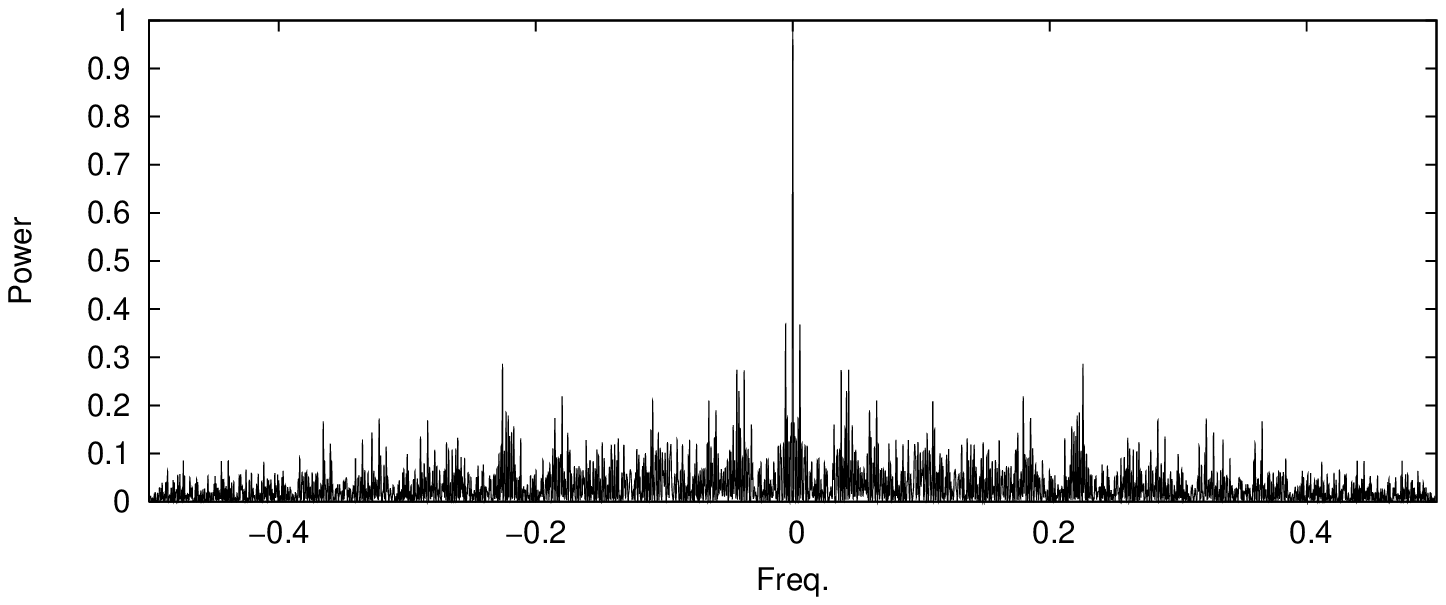}}
\resizebox{\hsize}{!}{\includegraphics{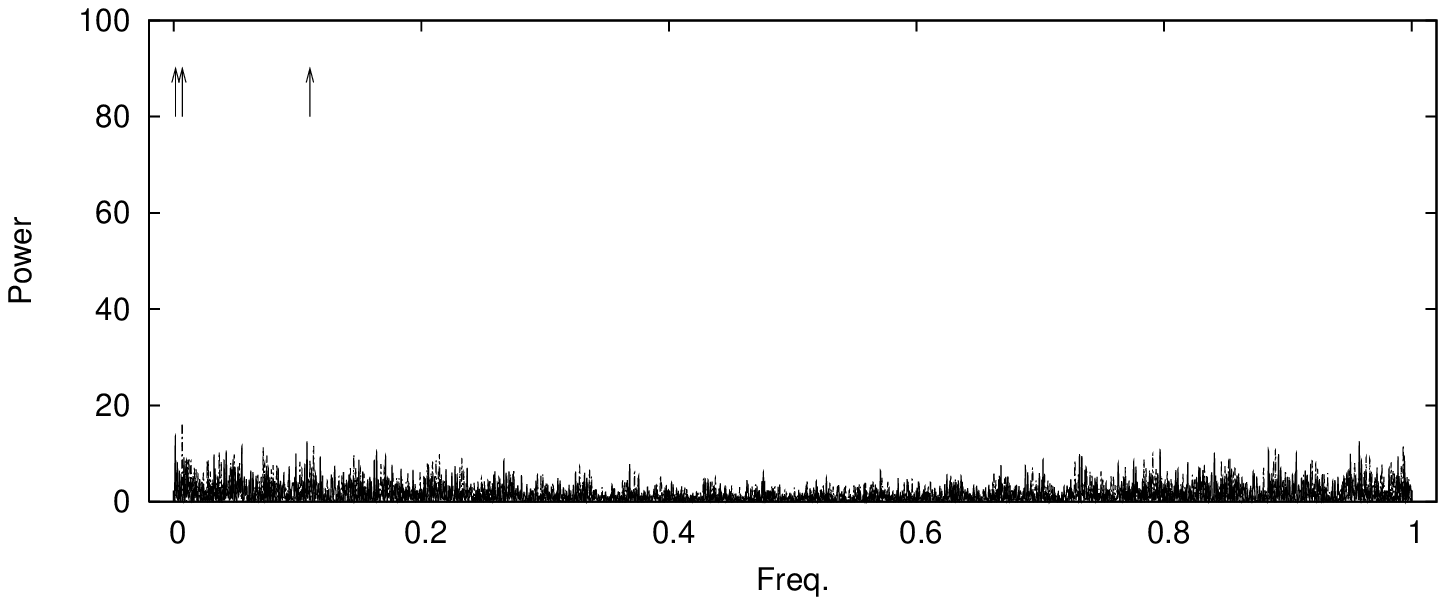}}
\caption{{\sl Upper panel:} Power spectrum of measured {\RV}s of
{\Halpha} and \ion{He}{i} 6678\,{\AA} in the frequency range from
$0.001\,{\cycday}$ to $1\,{\cycday}$,
{\sl middle panel:} spectral window, and
{\sl lower panel:} power spectrum of the prewhitened residual with
0.0014\,\cycday.
In all panels frequencies at the $x$-axis are given in {\cycday}.}
\label{hrpwr}
\end{figure}

\begin{table}[b]
\caption{Candidate frequencies resulting from the PDM period search.}
\label{pdmper}
\begin{tabular}{lrl}
\hline
frequency (\cycday) & period (days) & $\theta$ \\
\hline
0.0069 & 144.92753 & 0.3841 \\
0.0014 & 714.28571 & 0.4092 \\
0.1077 &   9.28505 & 0.4668 \\
\hline
\end{tabular}
\end{table}

\begin{figure}[t]
\resizebox{85mm}{!}{\includegraphics{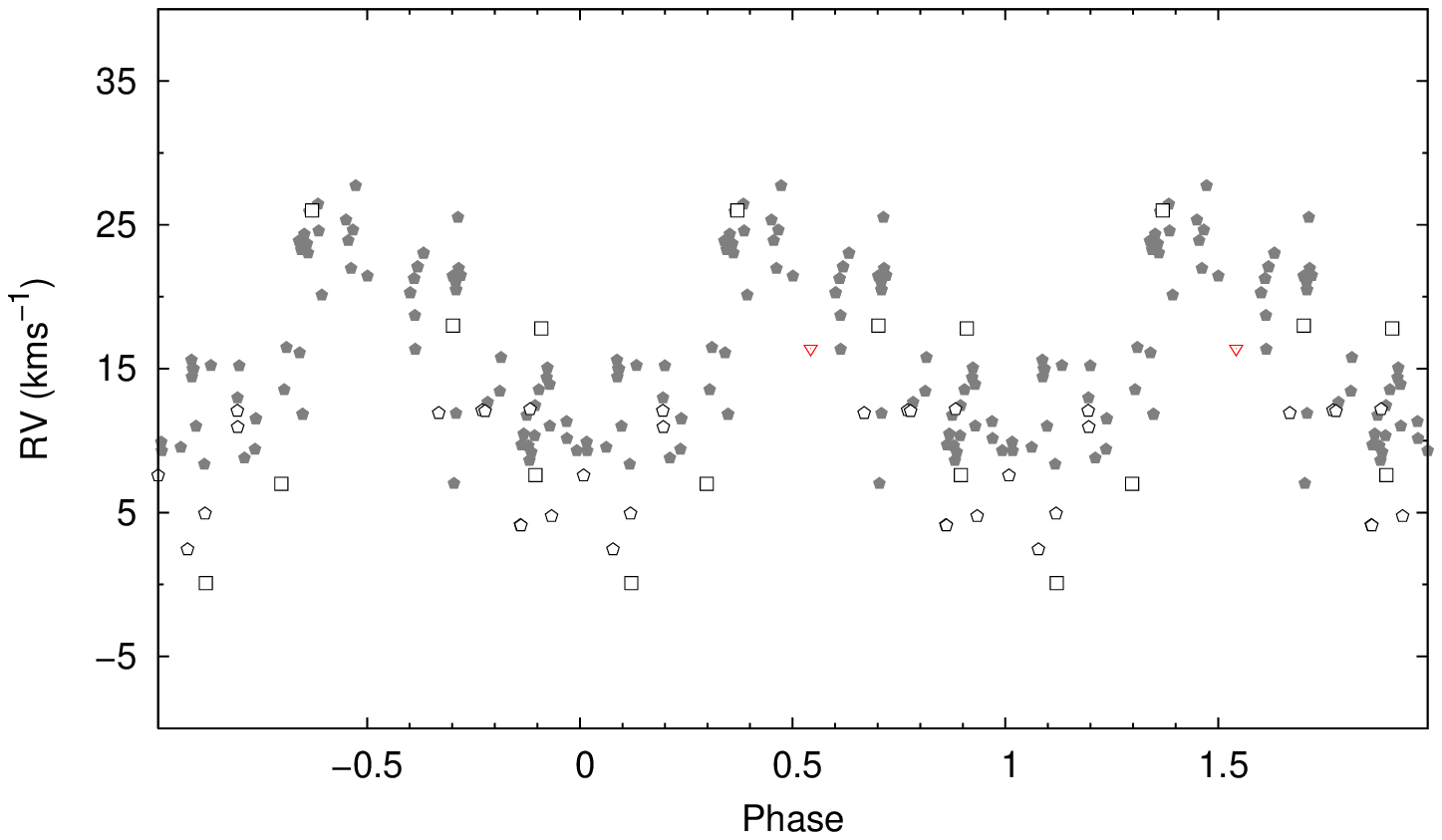}}
\resizebox{85mm}{!}{\includegraphics{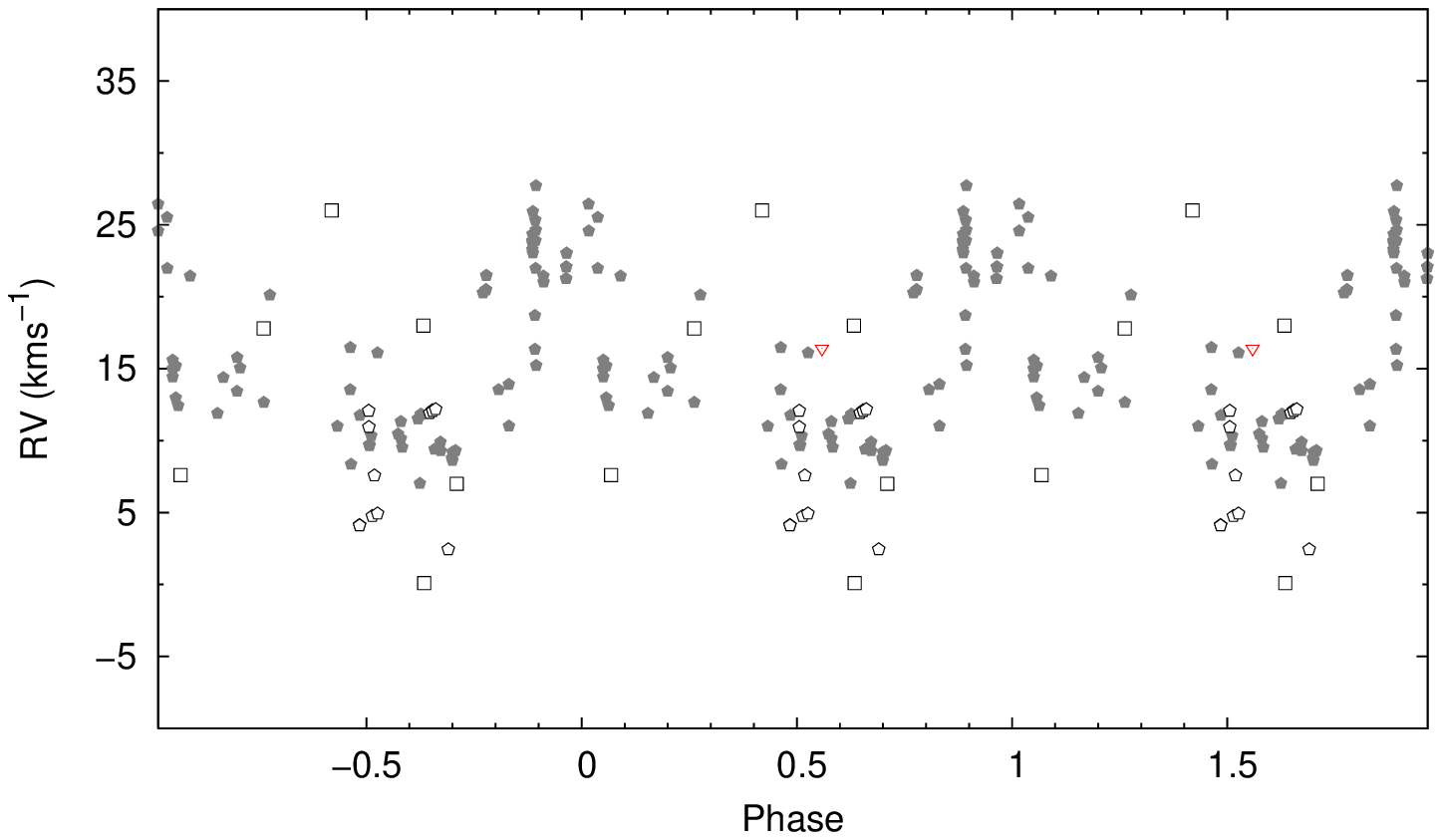}}
\resizebox{85mm}{!}{\includegraphics{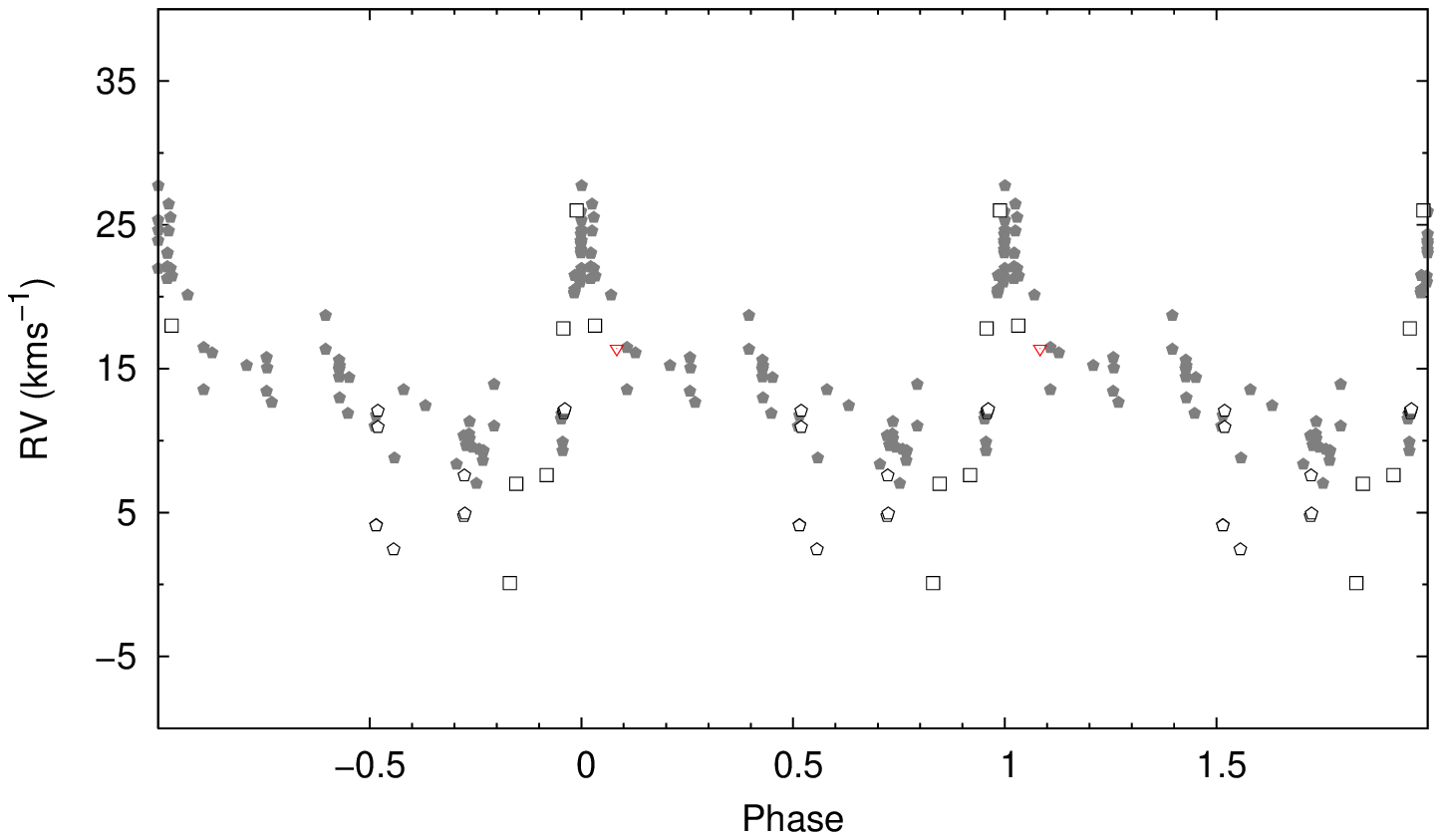}}
\caption {Phase diagram of measured {\RV}s for {\Halpha} lines folded
with different
frequencies
displayed from top to bottom,
$0.1077\,{\cycday}$ ($9\fd28505$), $0.00689\,{\cycday}$ ($144\fd9275$),
and $0.0014\,{\cycday}$ ($714\fd2857$).
Filled circles ($\bullet$) and open circles ($\circ$) denote data
obtained from our observations at Ond\v{r}ejov and Rozhen, respectively.
Open squares ($\square$) and open triangles ($\triangledown$) represent
data from \cite{DAObin3} and \cite{brazilie}, respectively.}
\label{pdm145}
\end{figure}

One of the most fundamental parameters of binary systems is the orbital
period.
The present {\RV}s span an interval of 1801 days.
From the inspection of the time plot one can conclude that
\begin{enumerate}
\item
the present {\RV} measurements of {\Halpha} and of some other metalic
lines show clear evidence of long-term cyclic variability, and
\item
the relatively large scatter between individual measurements on
successive nights indicates that shorter period variability could be
present in the data distribution.
\end{enumerate}

No period search of the {\RV} measurements has been carried out so far
for \hvea.
Improper data distribution, gaps, and timing of the observations are the
most common problems in the period search process.
The period search was carried out separately for several sets of
measurements, and two independent numerical period searching routines
were used.
The first one is based on the phase dispersion minimization (PDM)
technique \citep{Stell78}, the other one, {\PERIOD} \citep{Berg90}, is
based on a Fourier analysis.

For {\PERIOD}, starting values for the frequencies need to be given, and
frequency values are improved within the limits given by the window
function.
We used this program to search for periodicities in both {\Halpha} and
\ion{He}{i} 6678\,{\AA} {RV} data sets in the range between 1 to 1000
days, with an expected frequency resolution of $0.00055\,\cycday$.
The upper panel of Fig.\,\ref{hrpwr} displays the periodogram of the
detected frequencies in this period range.
It shows three candidate frequencies (indicated by arrows in the upper
panel of Fig.\,\ref{hrpwr}), namely 0.0069\,{\cycday},
0.0014\,{\cycday}, and 0.1077\,{\cycday}.
The frequency 0.0069\,{\cycday} has power 61, and is dominant.
The middle panel of Fig.~\ref{hrpwr} shows the spectral window.
In the lower panel of Fig.\,\ref{hrpwr} we show the power spectrum of
the residual periodogram prewhitened for 0.0014{\cycday} (using
{\PERIOD}), which clearly shows the disappearance of all peaks related
to all three candidate frequencies.

Using the PDM technique we searched for periodicity in intervals
$1\!-\!10$\,days, $1\!-\!100$\,days, and $1\!-\!1000$\,days.
The results are summarized in Table~\ref{pdmper} and agree with the ones
obtained with {\PERIOD}.
Figure~\ref{pdm145} illustrates the phase diagrams of measured {\RV}s
for the {\Halpha} line folded with the $0.1077\,{\cycday}$,
$0.0069\,{\cycday}$ and $0.0014\,{\cycday}$ frequencies.

Although the frequency 0.0069\,{\cycday} has the highest power (cf.
Fig.\,\ref{hrpwr} and Table\,\ref{pdmper}), we suspect that this
frequency is not a real one.
First, the historical observations of \cite{DAObin3} are out of phase by
about half of the period (open squares in the middle panel of
Fig.\,\ref{pdm145}).
Second, the corresponding {\RV} curve predicts a maximum of {\RV}s near
HJD~2453350, which was not confirmed by our observations.
Values around $\sim\!\!15{\kms}$ appear instead of maximum ones.
In addition, the interval of 145 days (or approximately five months)
roughly corresponds to the gap between the end of one season in
March/April and the next start of observations in August/September.
Third, the candidate frequency 0.0069 {\cycday} is almost exactly equal
to $5\times0.0014$\,{\cycday}.
The relative scatter of individual {\RV} values folded with the
0.1077\,{\cycday} frequency is quite large.
Following these arguments, we propose the frequency at 0.0014\,{\cycday}
as the dominant frequency.

\subsection{Orbital solution}

\begin{table*}[h]
\caption{Orbital elements of {\hvea} with {\FOTEL}.}
\label{fotres}
\begin{center}
\begin{tabular}{lrrrr} \hline
\hline
                     &                   &      &         & \\
\textbf{Element}     &
\textbf{Solution I}  &
\textbf{Solution II} &
\textbf{Solution III}&
\textbf{Solution IV} \\
                     &                   &      &         &   \\
\hline
P[d]           & $715.03\pm1.06$
               & $713.83\pm1.49$
               & $719.79\pm0.17$
               & $710.49\pm0.98$\\              
$T_\mathrm{periast.}$ & $51501.61\pm4.26$
                      & $51504.34\pm5.93$
                      & $51480.31\pm3.88$
                      & $51503.67\pm3.09$\\                
$K(\kms)$      & $7.89\pm0.53$
               & $7.58\pm0.55$
               & $7.73\pm0.44$
               & $8.23\pm0.23$  \\     
$e$            & $0.79\pm0.02$
               & $0.72\pm0.03$
               & $0.70\pm0.02$
               & $0.73\pm0.01$ \\             
\\
$\omega[\deg]$	& $306\pm5.87$
		& $310\pm6.99$
		& $304\pm7.03$
		& $286\pm3.70$\\             
\\
\\
$\gamma_{1}(\kms)$ &13.47& 13.39 & 13.44 & -- \\  
$\gamma_{2}(\kms)$ & --  & 9.54  &  9.88 & -- \\ 
$\gamma_{3}(\kms)$ & --  & --    & 10.63 & -- \\
$\gamma_{4}(\kms)$ & --  & --    & 10.52 & -- \\
$\gamma_{5}(\kms)$ & --  & --    & --    &13.23 \\
$\gamma_{6}(\kms)$ & --  & --    & --    &13.36 \\
$\gamma_{7}(\kms)$ & --  & --    & --    &10.97 \\
$\gamma_{8}(\kms)$ & --  & --    & --    &13.10 \\
                   &     &       &       &      \\
$f(m)$            & $0.0095$
                  & $0.0104$   
                  & $0.0126$
                  & $0.0132$ \\      
& \\ \hline
No. of RVs             & $64$  
                       & $76$ 
                       & $83$ 
                       & $246$  \\           
\\ \hline
$rms$ (\kms)           & $1.76$ 
                       & $2.08$ 
                       & $2.22$ 
                       & $1.94$ \\          
\\
\hline
\end{tabular}   
\end{center}
{\em Note:}
The velocity $\gamma_1$ is based on Ond\v{r}ejov data, $\gamma_2$ on
Rozhen data, $\gamma_3$ on OPD data \citep{brazilie}, and $\gamma_4$ on
historical data of \cite{DAObin3}.
In solution IV $\gamma_5$, $\gamma_6$, $\gamma_7$, and $\gamma_8$ stand
for {\Halpha}, {\hei} 6678\,{\AA}, {\siii} 6347\,{\AA}, and {\siii}
6371\,{\AA}, respectively.
\end{table*}

\begin{figure}[t]
\resizebox{\hsize}{!}{\includegraphics{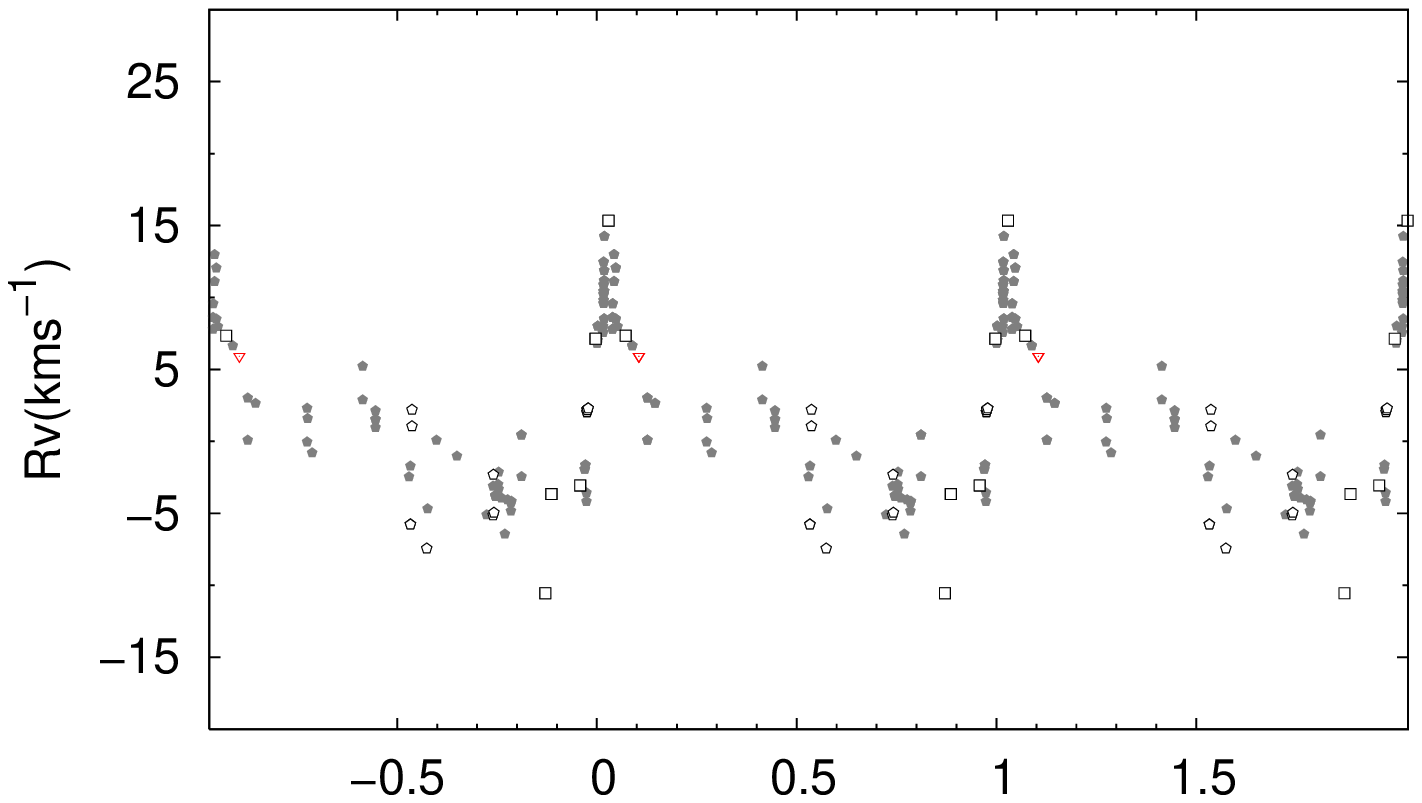}}
\resizebox{\hsize}{!}{\includegraphics{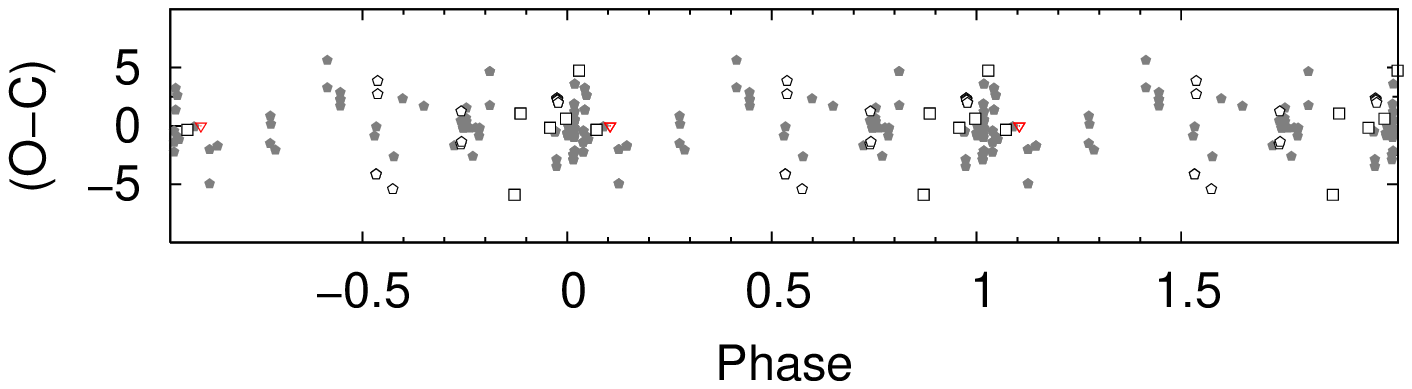}}
\caption {{\it Upper panel}: Radial velocity curve of the primary
component corresponding to the {\FOTEL} orbital solution III
($P=719\fd79$), which was based on {\Halpha} {\RV} measurements of
Ond\v{r}ejov, Rozhen, and OPD spectra and on {\RV} values from
\cite{DAObin3}.
{\RV} values are displayed in the centre-of-mass reference system.
{\it Lower panel}: Phase plot of $O-C$ deviation from the {\RV} curve
for Solution~III.}
\label{fotiii}
\end{figure}

The variability of the radial velocities with frequency
0.0014\,{\cycday} can most likely be explained by binarity, and it
supports the earlier idea of \cite{DAObin3} that the star is a
spectroscopic binary.

Starting from the frequency 0.0014\,{\cycday}, we performed 4 different
orbital solutions using the program {\FOTEL} developed by
\citet{fotelorig, fotelman}.
For the solutions denoted as I, II, and III (see Table~\ref{fotres}),
the orbital elements were derived for {\Halpha} radial velocity
variations with data observed from different sites.
Solution I is only performed with Ond\v{r}ejov data, Solution II is
obtained using {\Halpha} measurements from both Ond\v{r}ejov and Rozhen
observations, Solution III is obtained using all of the available data,
namely Ond\v{r}ejov and Rozhen observations, one spectrum from OPD, and
including the historical measurements.
Solution IV was obtained using radial velocities of all available
spectral lines in the {\Halpha} region ({\Halpha}, {\hei} 6678\,{\AA},
{\siii} 6347\,{\AA}, and {\siii} 6371\,{\AA}).

In our calculations we allowed the period, eccentricity, periastron
longitude, and semiamplitude to converge.
For each data set {\FOTEL} also allows individual $\gamma$-velocities to
be determined.
Figure~\ref{fotiii} illustrates the {\FOTEL} Solution III with its $O-C$
residuals.
We searched for additional periodicities in the  residuals but did not
find any significant periods.

We adopt the Solution III ($P=719\fd79\pm0\fd17$, $e=0.70\pm0.02$) as an
orbital solution of {\hvea}.

\section{Component B}

\begin{figure}[t]
\resizebox{\hsize}{!}{\includegraphics{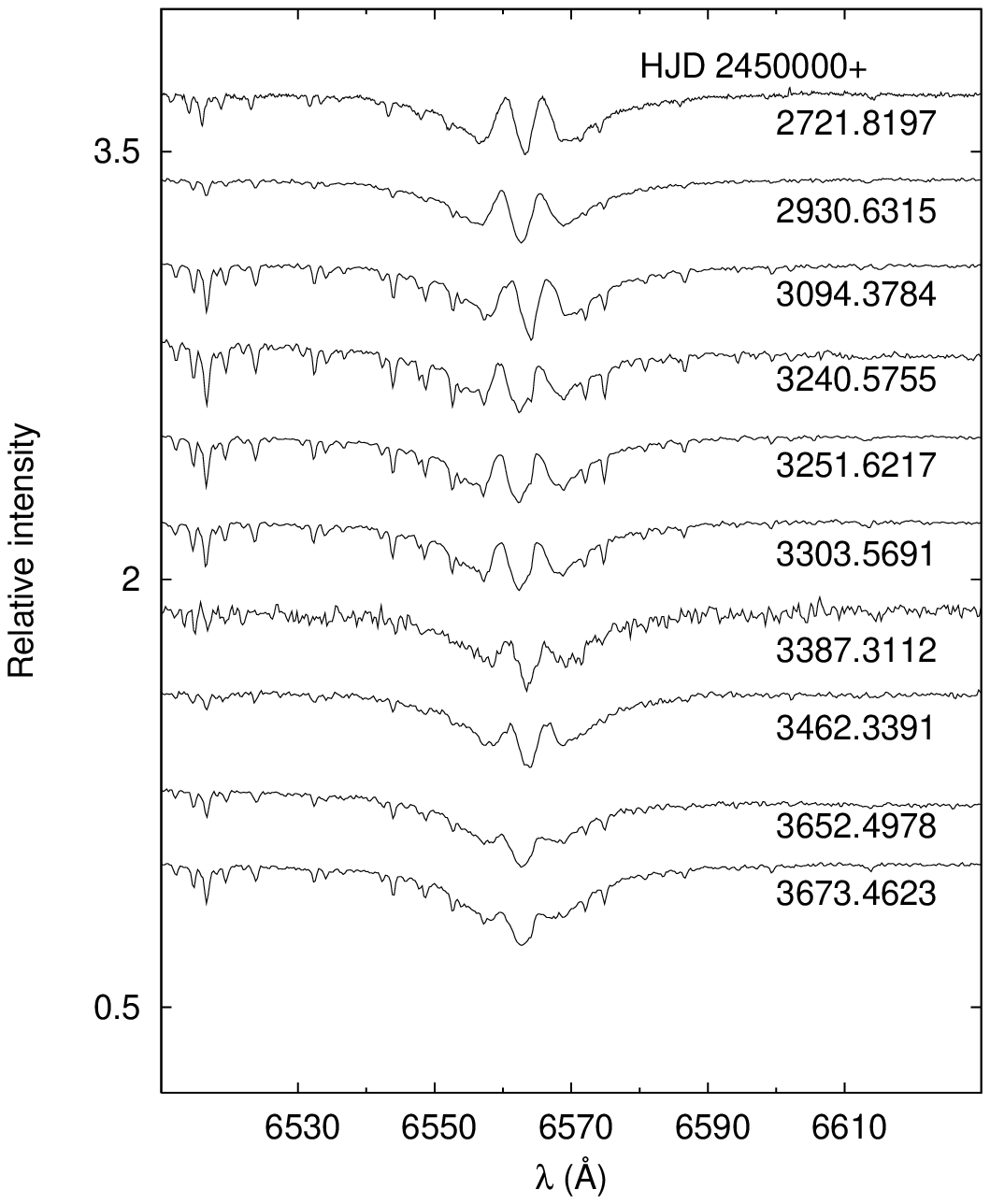}}
\caption{The evolution of the {\Halpha} profile of {\hveb}.}
\label{halpB}
\end{figure}

The B component is a Be star with a relatively weak emission in
{\Halpha} and almost negligible one in {\Hbeta}.
Since 2003, the weak emission in {\Halpha} further weakened, indicating
an approaching end of the Be phase.
No time scale
may be given for this apparently long-term variability,
since only 4 years of observations are available.
Long term variability (strengthening and weakening of emission) is
typical of Be stars on time scales from years to tens of years
\citep[for a review see][]{amhsapp}.
For example, $\varkappa$~Dra has quite a well-established long term time
scale, and its most recent determination is $22.11$ years \cite[see][and
references therein]{brnokap}.
The {\Halpha} profiles obtained by our observations are plotted in
Figure~\ref{halpB}.
We fitted the spectrum of {\compb} using the same methods as for
{\compa}.
We found $\Teff=(12500\pm500)\Kelvin$, $\log g=(3.5\pm0.5)$, and $\varv
\sin i=(230\pm5)\kms$.

\subsection{Radial velocities}
\label{secRVB}

\begin{figure}[b]
\resizebox{\hsize}{!}{\includegraphics{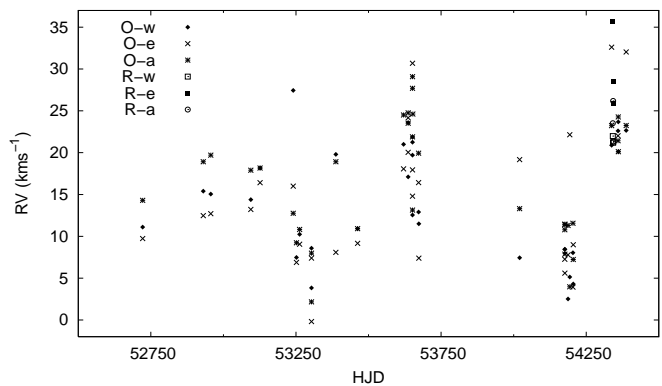}}
\caption{Time evolution of the {\Halpha} radial velocities of {\hveb}.
Measurements of line wings are denoted by {\tt w}, measurements of
emission are denoted by {\tt e}, and measurement of the central
absorption are denoted by {\tt a}.
Capital letters denote the observatory, {\tt R} stands for Rozhen,
{\tt O} stands for Ond\v{r}ejov.}
\label{rvhalpB}
\end{figure}

Owing
to the high rotational broadening of lines in the {\compb} spectrum, the
helium and silicon lines, which were used for {\RV} measurement of
{\compa}, are too shallow in {\compb}, and therefore unusable for {\RV}
measurement.
Consequently, we measured the RVs of {\compb} only from the {\Halpha}
line.
{\RV}s were measured with the profile inversion technique.
However, the rapid rotation of the {\compb} and relatively low S/N made
it impossibile to arrive at precise RV values.
Cross-correlation techniques, such as least-square deconvolution
\citep[LSD,][]{d97} would not improve the RV accuracy as too few line
profiles are available in our spectra.
Also, the bisector method does not give more accurate results because of
the limited quality and not high enough S/N of our spectra.

The {\Halpha} line in the B component is characterized by a double-peak
emission with peaks of equal strength.
We measured the {\RV}s of three different features at the {\Halpha}
line, namely of the line wings, of the emission part, and of the central
absorption.
The results are listed in Tables~\ref{rvbond} and \ref{rvbroz} and
plotted in Fig.\,\ref{rvhalpB}.

Although the {\RV}s are evidently variable, a period search did not give
any reasonable result.
That is why we cannot make any conclusions about the possible binarity
of {\hveb} until a more complete set of more accurate data has been
obtained.

\section{The possible A+B system}

There is a question as to whether the components of this visual pair are
physically bound.
However, there is not enough data to help to resolve this question.
The history of all relative position measurements of {\hve} in the sky
(see Section~\ref{relpos}) do not allow any conclusion about their
relative motion.
Slightly different values can be attributed to errors in individual
measurements.

Unfortunately, only a common parallax measurement from {\Hipparcos} is
available.
Since individual measurements are missing, we cannot easily say whether
the A and B components are at the same distance.
However, study of visual binaries by \cite{vis82} suggests that {\hve}
could be a physical system (with a period shorter than 1Myr), since both
components have similar proper motion and corresponding photometric
parallaxes, which agree with the hypothetical parallax.

In case the visual components of {\hve} are physically bound, their
similar spectral type suggests that they are of similar age.
Then there is the question of why one of these stars developed an
envelope that causes {\Halpha} emission and the other does not.
The emission could not be caused by mutual interaction between
components A and B, simply because the distance between the components
is too large.
A different mechanism for the origin of the Be phenomenon in {\hveb} has
to be found \cite[see, e.g., the review by][]{classbe}.
The emission in the {\compb} is most probably connected with its rapid
rotation.
Projected rotational velocity of {\compa} is lower, and this star also
has no emission.
On the other hand, since the inclination angles are unknown, we can not
exclude the possibility that the {\compa} is also rapidly rotating, as
there is apparently no correlation between directions of spin axes for
wide binaries \citep[][]{vsinicorr}.

\subsection{Comparison with other visual binaries}

{\hve} is not the only example of a visual multiple system with similar
components, with some of them also showing the Be phenomenon.
The visual triple system $\beta$~Mon consists of three Be stars
\citep[see][and references therein]{betmonarg}, and all these three
stars are rapidly rotating \citep{ALe02}.
The projected rotational velocities of $\beta$~Mon~A and $\beta$~Mon~C
was reported as 260{\kms} and 250{\kms}, respectively, while
$\beta$~Mon~B is rotating more slowly (140\kms) and, interestingly, also
has weakest emission \citep{cogu73}.
This supports the connection between rapid rotation and the Be
phenomenon.
More visual binaries with Be-components may be found in \cite{Bebin}.

Investigation of such multiple systems with resolved visual components
(like {\hve} or $\beta$~Mon) may help when studying multiple partially
resolved or unresolved systems with Be stars like $o$~And
\citep[see][and references therein]{oandbeo} or $\beta$~Cep
\citep[see][and references therein]{betcepangl}, where light from all
stars is mixed, making the analysis extremely difficult.
The physical parameters and system characteristics derived from resolved
systems will put constraints on stellar structure and binary evolution
models, such that the unresolved systems can be modelled in a better way
with the improved models.
Careful study of resolved systems and careful determination of their
parameters are very important feedback on models.

\section{Conclusions}

\begin{table}
\caption{Summary of parameters of both components of {\hve}.}
\begin{tabular}{c|cc}
\hline
& \hvea & \hveb \\
\hline
$\alpha(\mathrm{J}2000)$ & $05\,32\,14.14$ & $05\,32\,14.56$ \\
$\delta(\mathrm{J}2000)$ & $+17\,03\,29.3$ & $+17\,03\,21.8$ \\
$V$ & 6.09 & 6.51 \\
\hline
\Teff & $12500\pm500$ & $12500\pm500$ \\
$\log g$ & $3.5\pm0.5$ & $3.5\pm0.5$ \\
$v\sin i$ (\kms) & $40\pm3$ & $230\pm5$ \\
Be star & no & yes \\
binary & SB1$^\star$ & SB1? \\
\hline
\end{tabular}
\\ $\star$ -- Binary parameters are listed in Table \ref{fotres}
\end{table}

This paper is the first attempt to collect all available information
about the stars in the visual binary system {\hve} and to correct errors
that appear in the SIMBAD database and Bright Star Catalogue.
We have presented results of our spectroscopic analysis.
{\hvea} \cite[$V=6.09$,][]{vis82} is a B type single-lined, eccentric,
($e=0.70\pm0.02$), spectroscopic binary with a period of
$719.79\pm0.17$\,days.
Model atmosphere analysis of this star yielded $\Teff=12500\Kelvin$,
$\log g=3.5$, and $\vsinia=40\kms$.
{\hveb} \cite[$V=6.51$,][]{vis82} is a Be star with variable radial
velocities, however, no reasonable orbital solution could be found based
on the available data.
Model atmosphere analysis of this star yielded the same effective
temperature and surface gravity as for the A~component, but with a
rotational velocity of $\vsinib=230\kms$.
A decrease in the {\Halpha} emission was recorded during the last 4.5
years of observations.
In addition, positions of X-ray (1RXS~J053214.9+170319, see
Section~\ref{xpaprsky}) and infrared (X0501+589, see
Section~\ref{infrak}) sources indicate that {\hveb} is more likely an
X-ray and IR source than {\hvea}.

Several other issues, such as the short term variability of {\hvea} or
the reliable period determination of {\hveb}, could not be addressed by
this paper owing to the lack of necessary data.
Consequently, further long-term observations of these stars are
desirable.
Future work should also include a detailed abundance analysis using
high S/N spectra and NLTE model atmospheres.
 
\begin{acknowledgements}
The authors would like to devote this paper to the memory of Dr. Izold
Pustylnik, with whom they consulted for the history of Struve's
observations in Tartu.
The authors would also like to thanks Zden\v{e}k Jan\'ak, Jan Elner, and
Petr \v{S}va\v{r}\'{\i}\v{c}ek for their help in early stages of the
work.
This research made use of the Washington Double Star Catalog maintained
at the U.S. Naval Observatory.
This research has made use of the NASA's Astrophysics Data System
Abstract Service.
Our work was supported by a grant of the Grant Agency of the Czech
Republic 205/08/0003.
The Astronomical Institute Ond\v{r}ejov is supported by project
AV0\,Z10030501.
\end{acknowledgements}

\newcommand{\Alicante}[1]{in The Be Phenomenon in Early Type Stars, IAU
	Coll. 175, M. A. Smith, H. F. Henrichs, \& J. Fabregat eds., ASP
	Conf. Ser. Vol. 214, p. #1}

\clearpage

\appendix

\section{Visual spectra of {\hvea} and {\hveb}.}

\begin{figure*}
\includegraphics[width=\hsize]{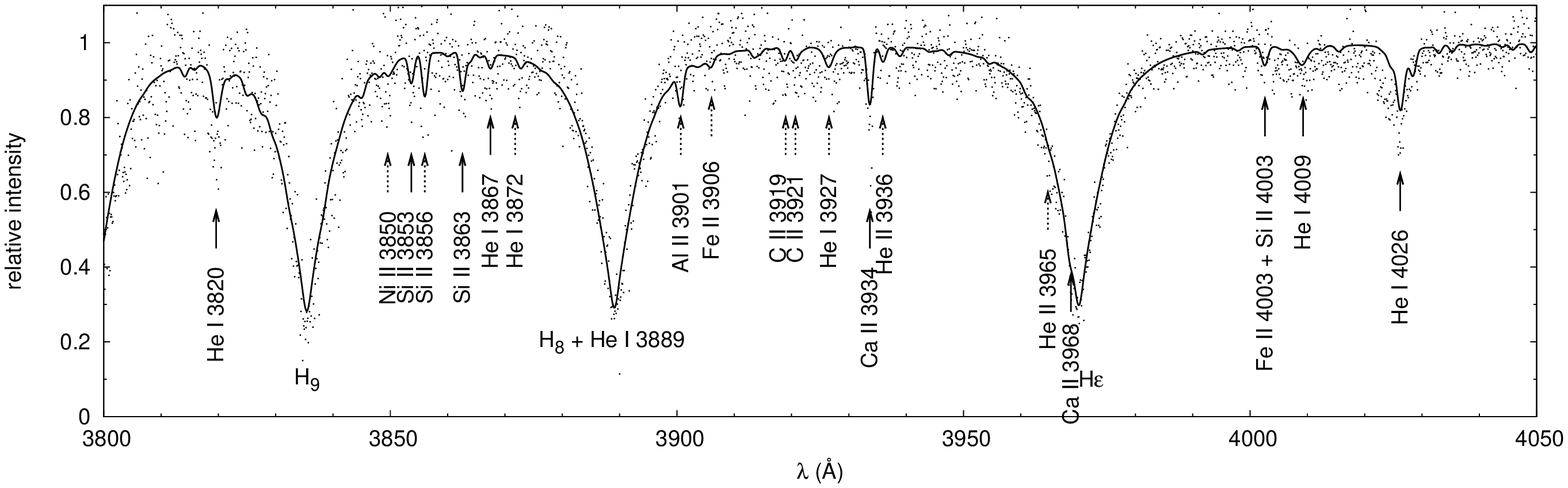}
\includegraphics[width=\hsize]{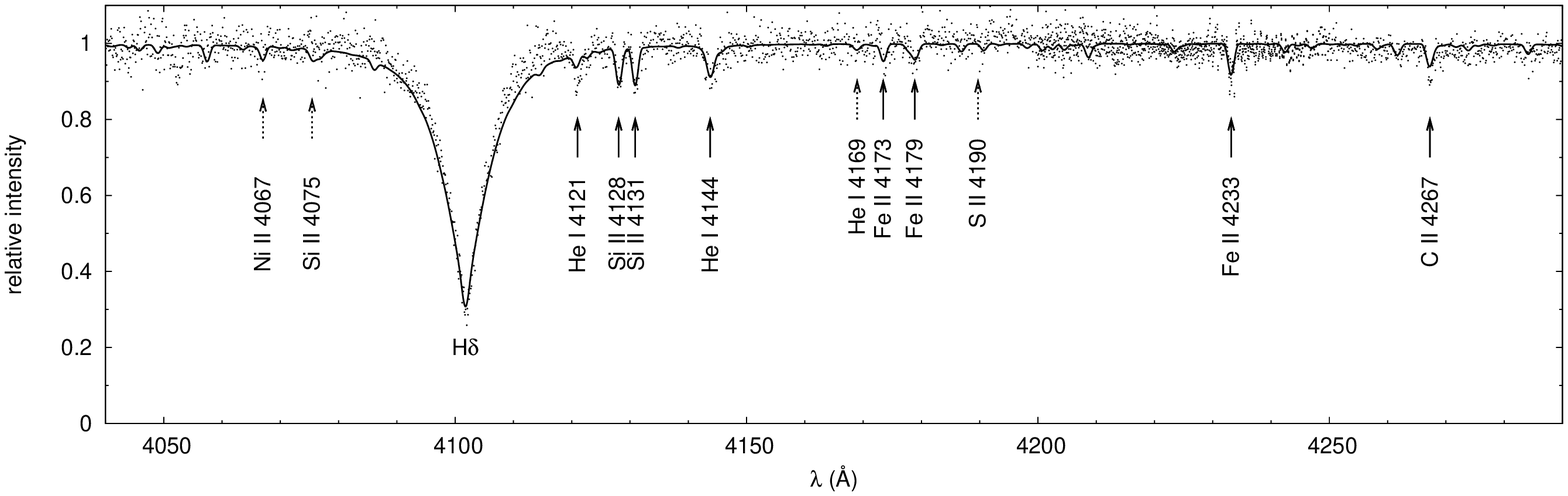}
\includegraphics[width=\hsize]{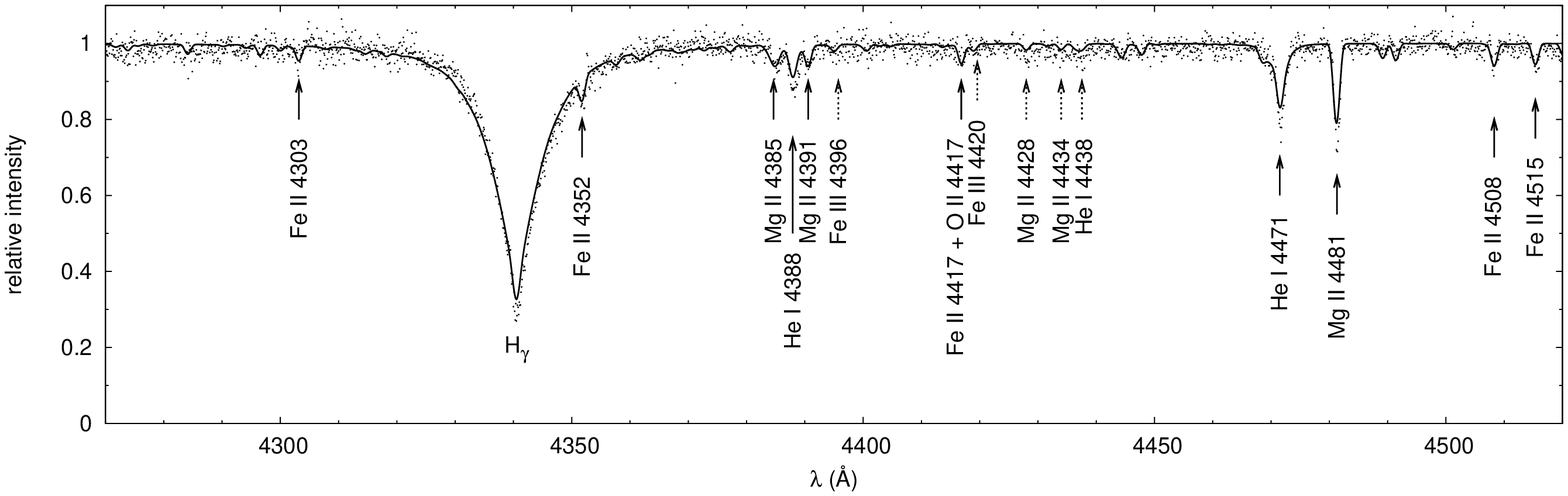}
\includegraphics[width=\hsize]{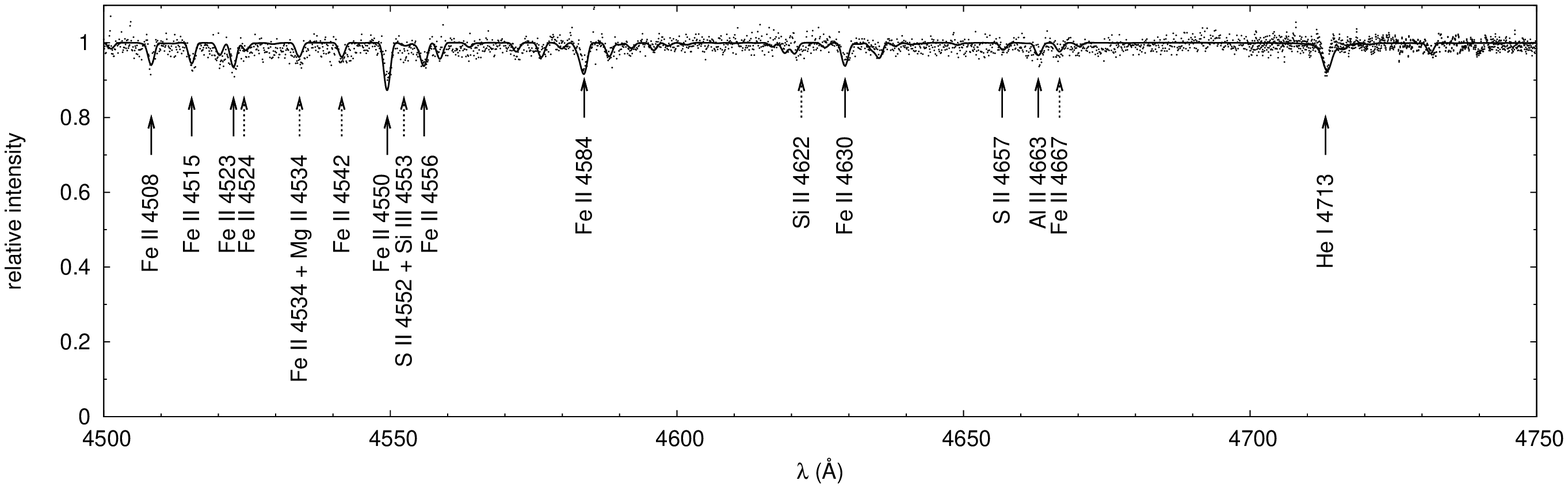}
\caption{Line identification of the spectrum of {\hvea} obtained with
the blue channel of the {\HEROS} spectrograph (dots) and comparison with
the synthetic spectrum calculated from the \cite{Kur13} LTE model
atmosphere $\Teff=12\,500\Kelvin$, $\log g=3.5$, rotationally broadened
with $\vsinia = 40\kms$ (full line).}
\label{HERAblue}
\end{figure*}

\addtocounter{figure}{-1}
\begin{figure*}
\includegraphics[width=\hsize]{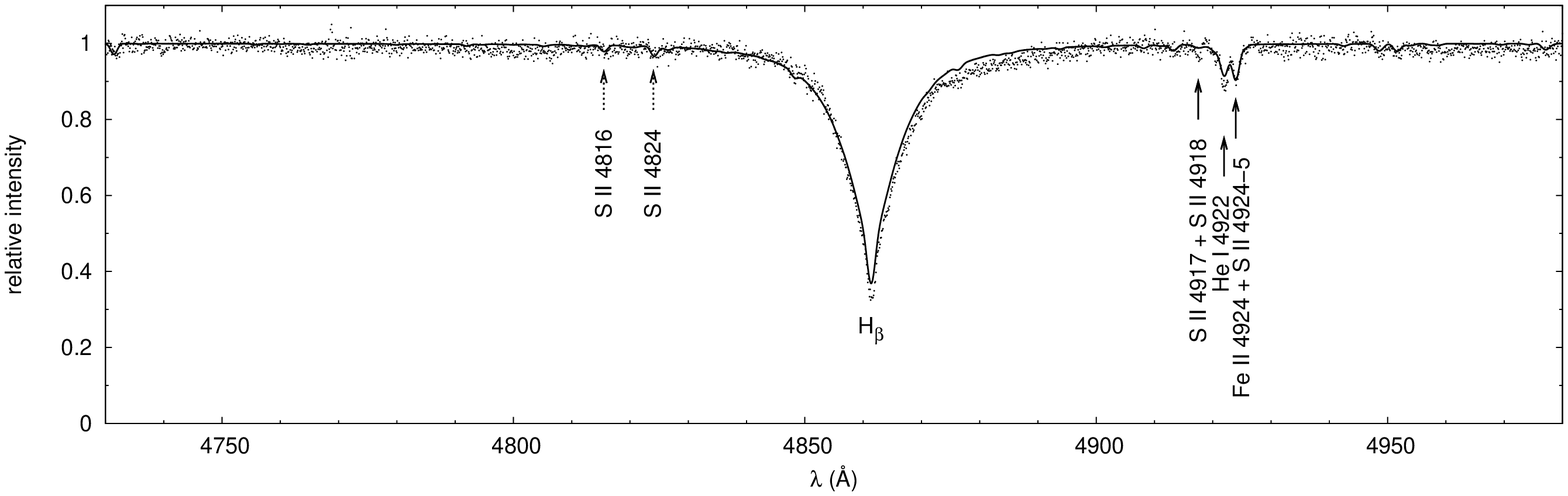}
\includegraphics[width=\hsize]{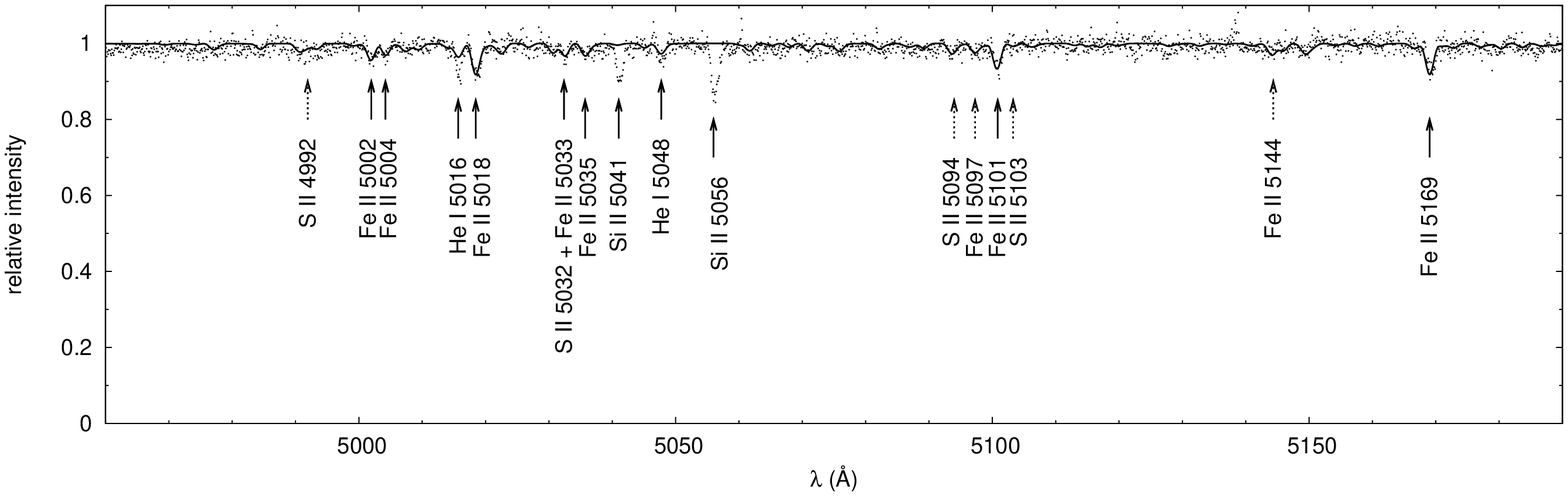}
\includegraphics[width=\hsize]{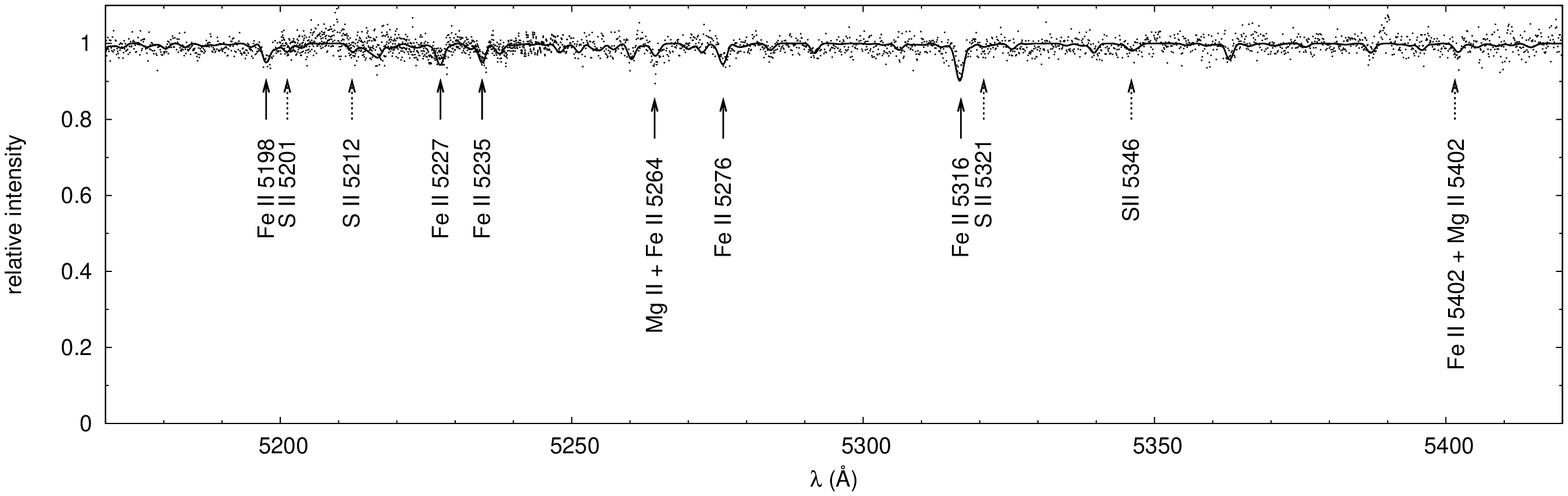}
\includegraphics[width=\hsize]{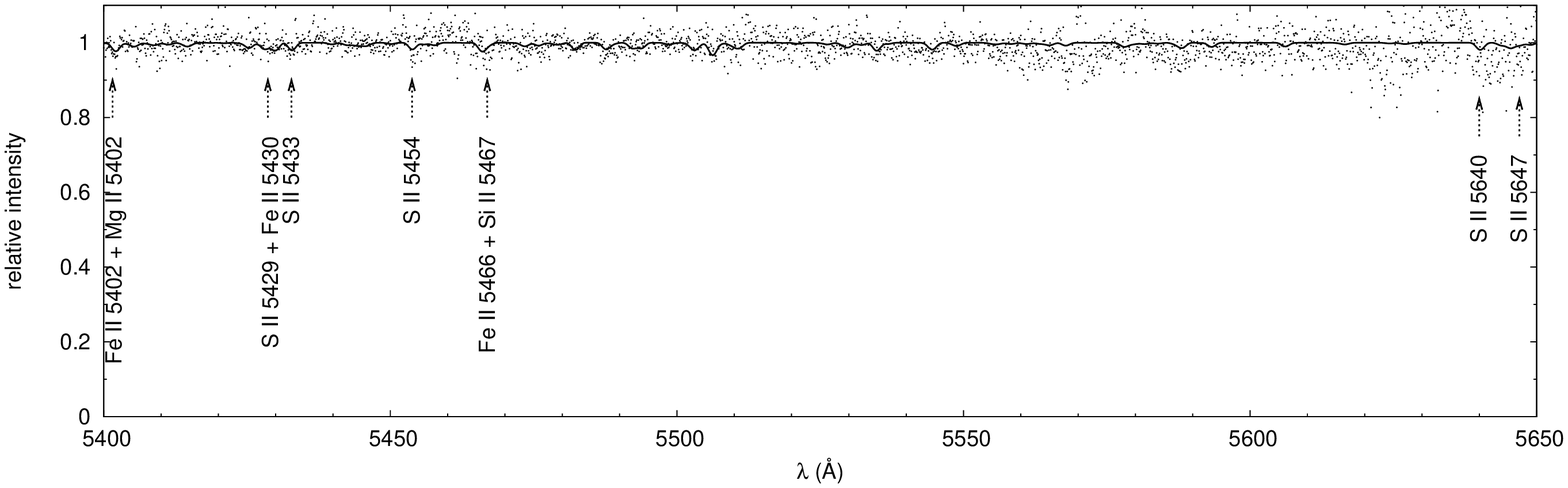}
\caption{contd.}
\end{figure*}

\begin{figure*}
\includegraphics[width=\hsize]{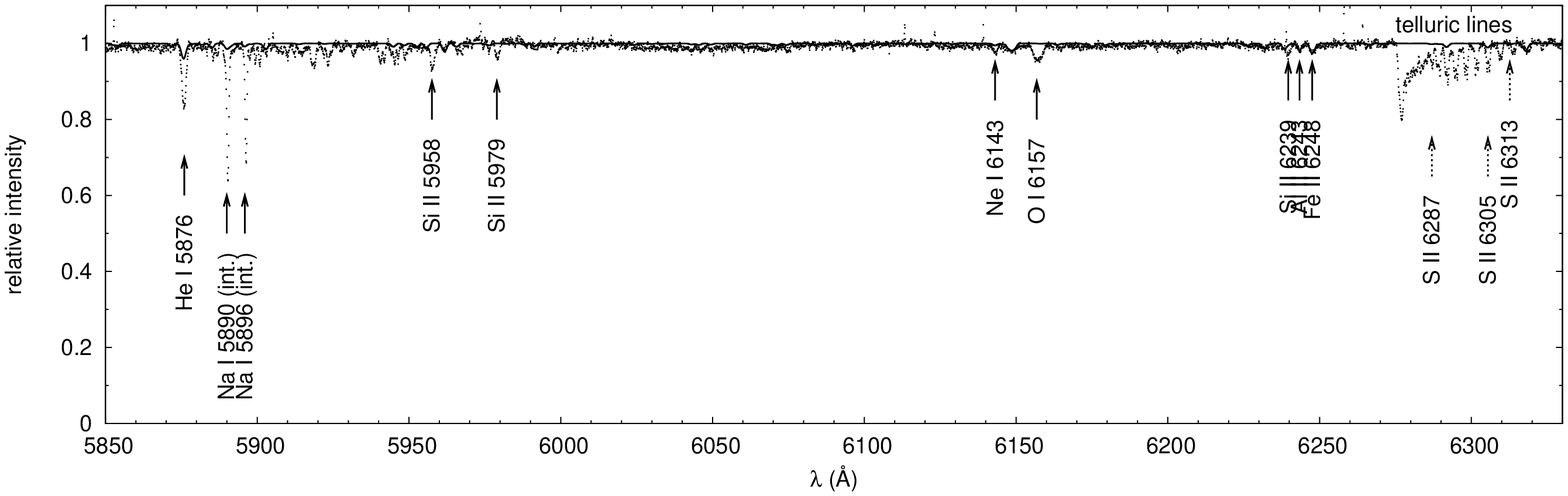}
\includegraphics[width=\hsize]{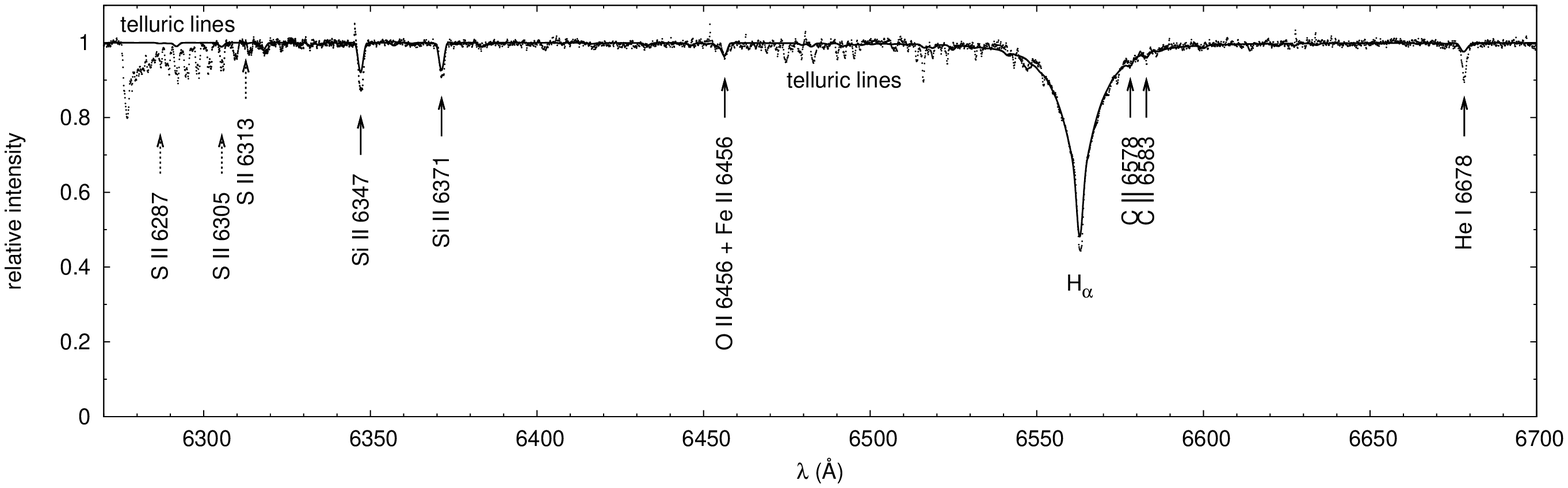}
\includegraphics[width=\hsize]{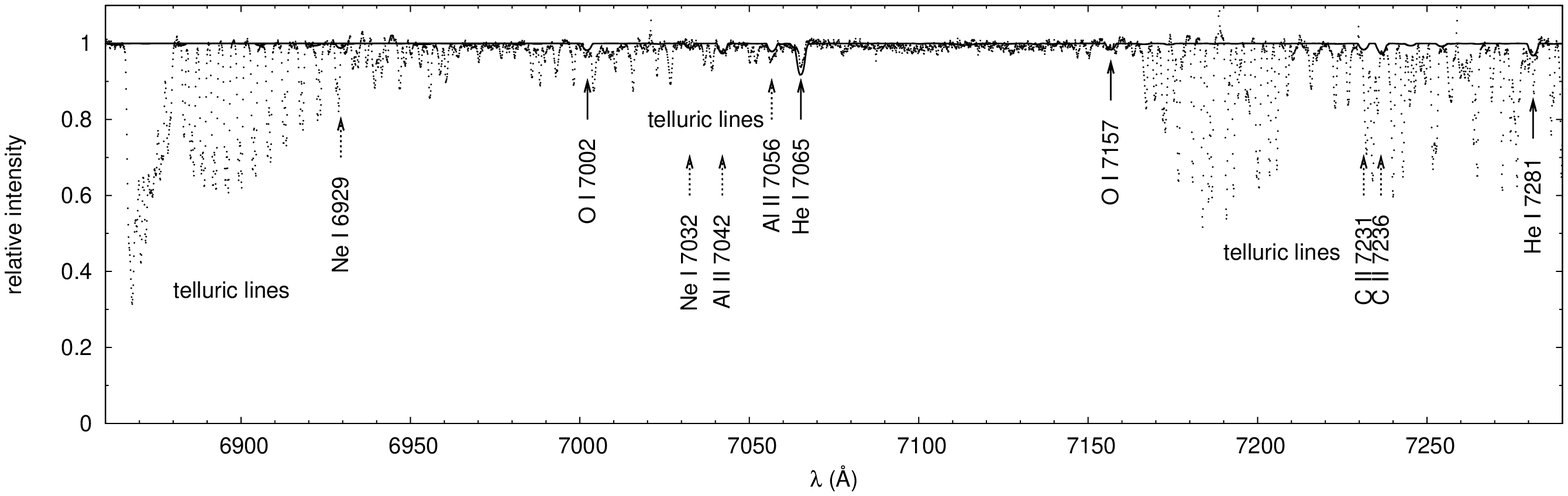}
\includegraphics[width=\hsize]{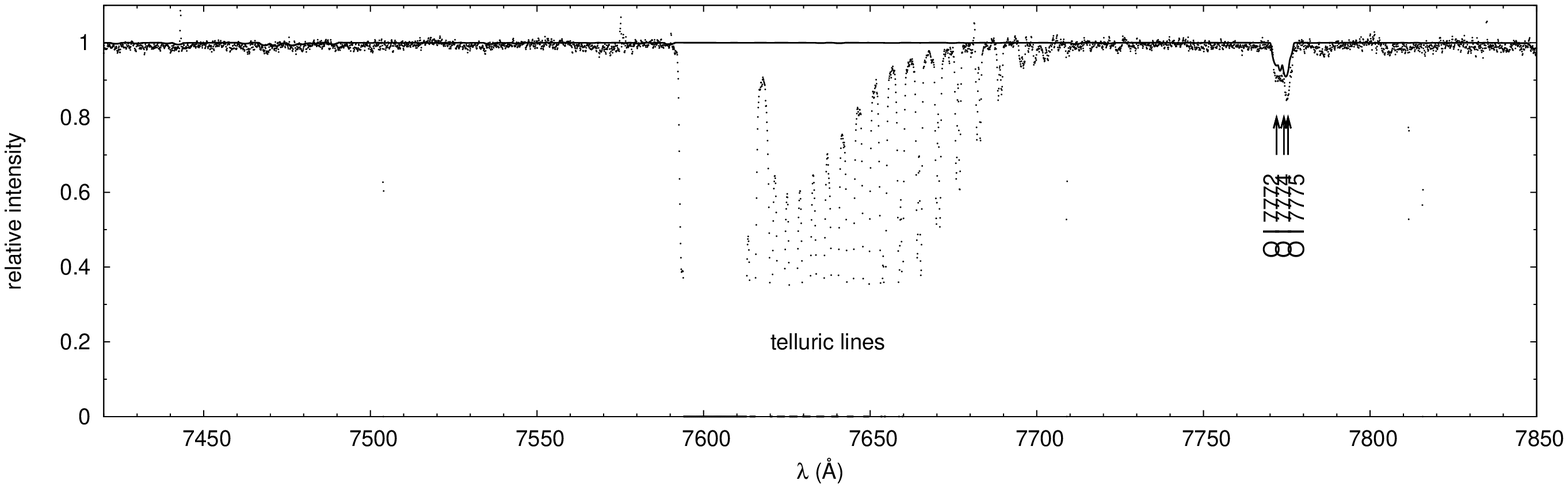}
\caption{Line identification of the spectrum of {\hvea} obtained with
the red channel of the {\HEROS} spectrograph (dots) and comparison with
the synthetic spectrum calculated from the \cite{Kur13} LTE model
atmosphere $\Teff=12\,500\Kelvin$, $\log g=3.5$, rotationally broadened
with $\vsinia = 40\kms$ (full line).
Note the presence of telluric lines and bands near $6280${\AA},
{\Halpha}, $6870${\AA}, $7250${\AA}, and $7590${\AA.}}
\label{HERAred}
\end{figure*}

\begin{figure*}
\includegraphics[width=\hsize]{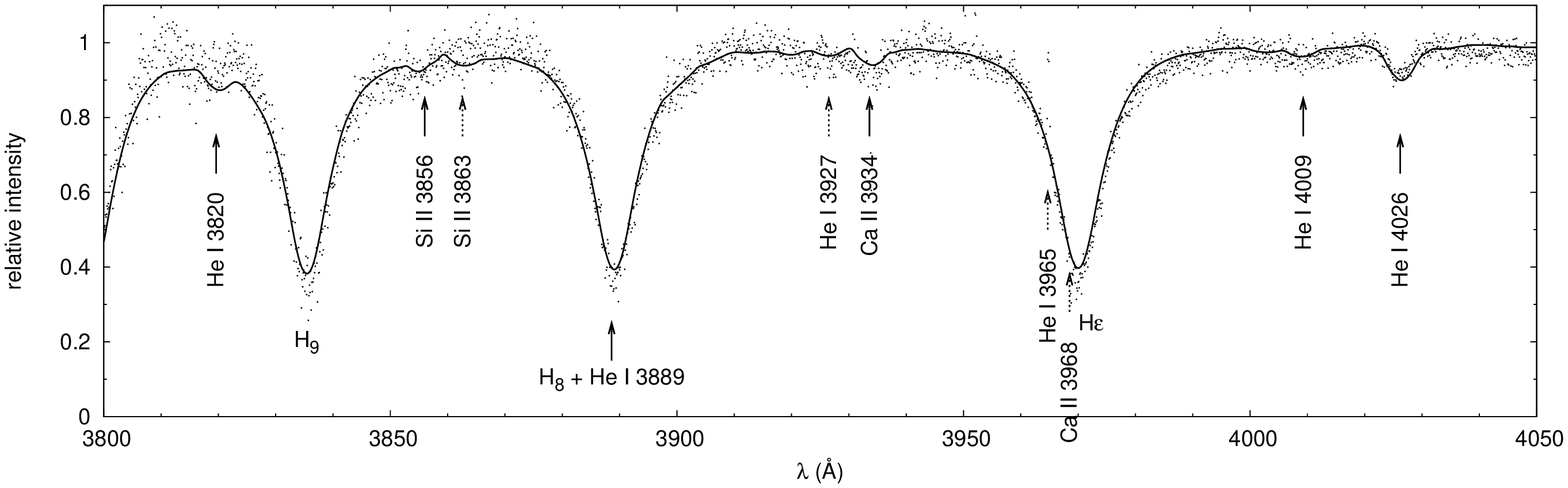}
\includegraphics[width=\hsize]{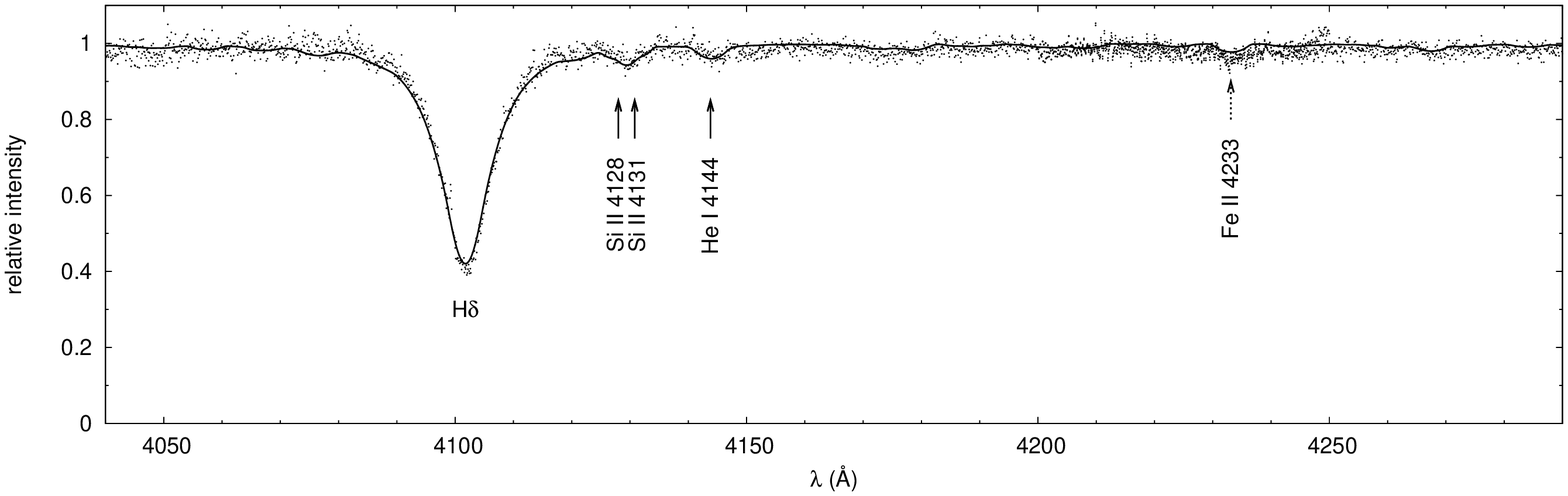}
\includegraphics[width=\hsize]{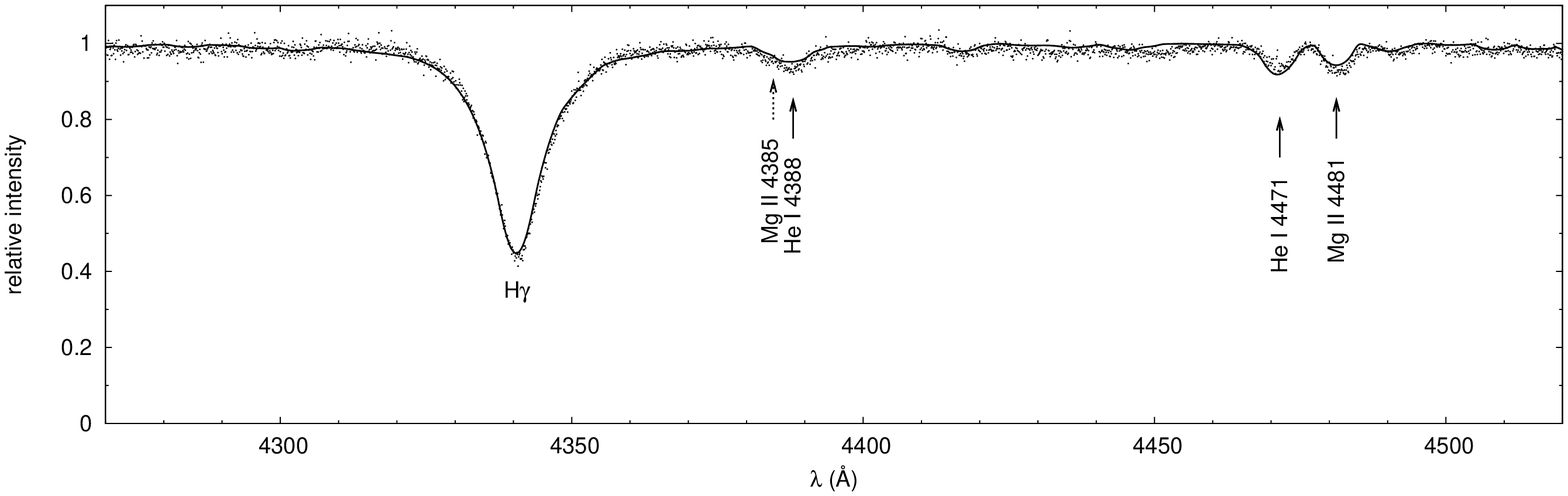}
\includegraphics[width=\hsize]{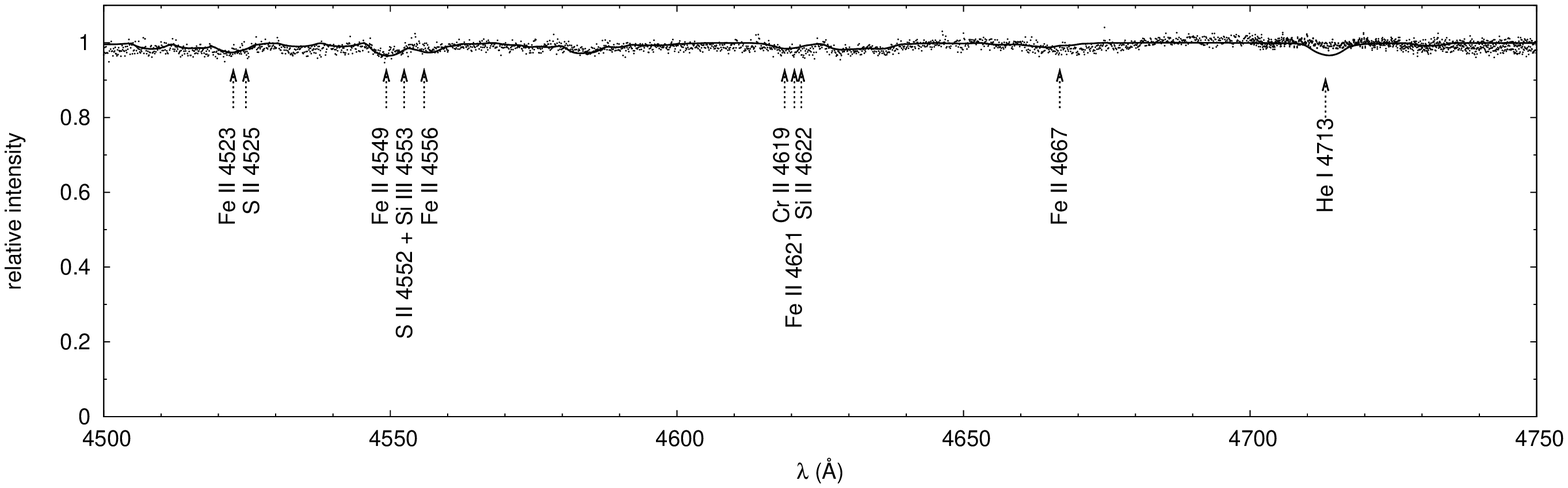}
\caption{Line identification of the spectrum of {\hveb} obtained with
the blue channel of the {\HEROS} spectrograph (dots) and comparison with
a synthetic spectrum calculated from the \cite{Kur13} LTE model
atmosphere $\Teff=12\,500\Kelvin$, $\log g=3.5$, rotationally broadened
with $\vsinib = 230\kms$ (full line).}
\label{HERBblue}
\end{figure*}

\addtocounter{figure}{-1}
\begin{figure*}
\includegraphics[width=\hsize]{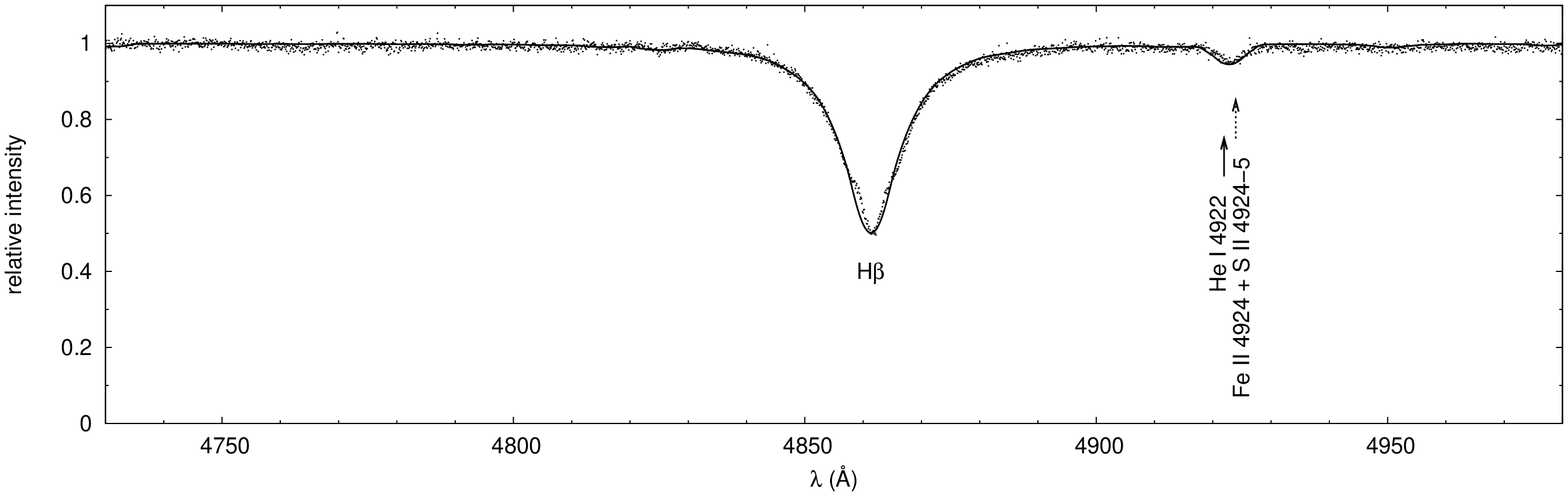}
\includegraphics[width=\hsize]{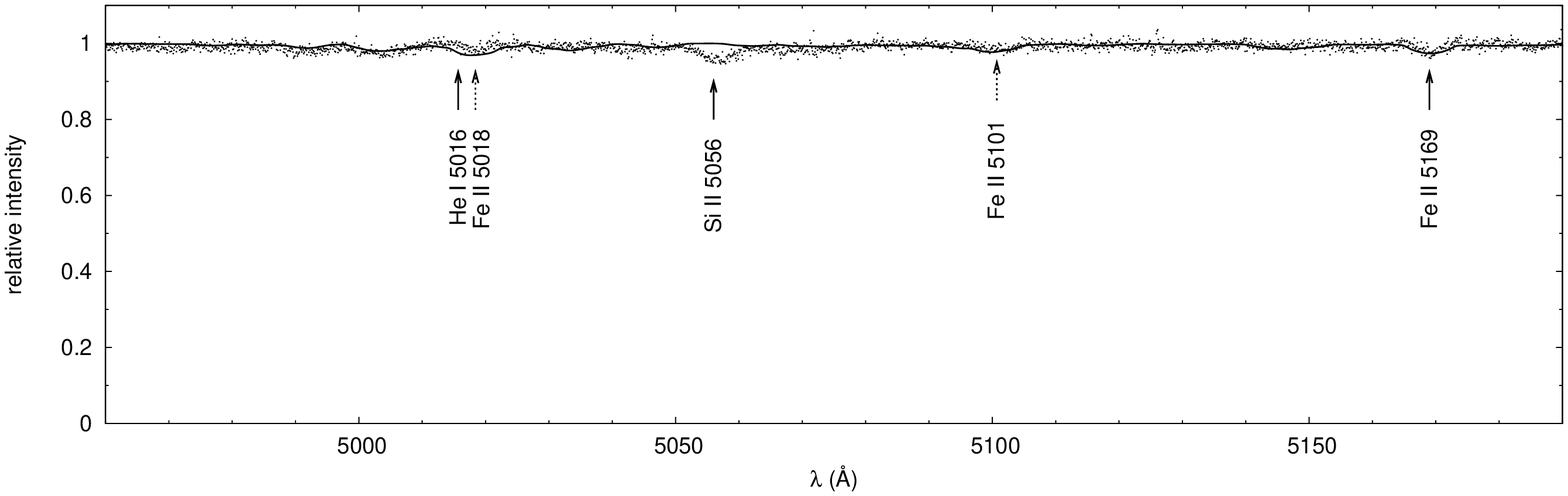}
\includegraphics[width=\hsize]{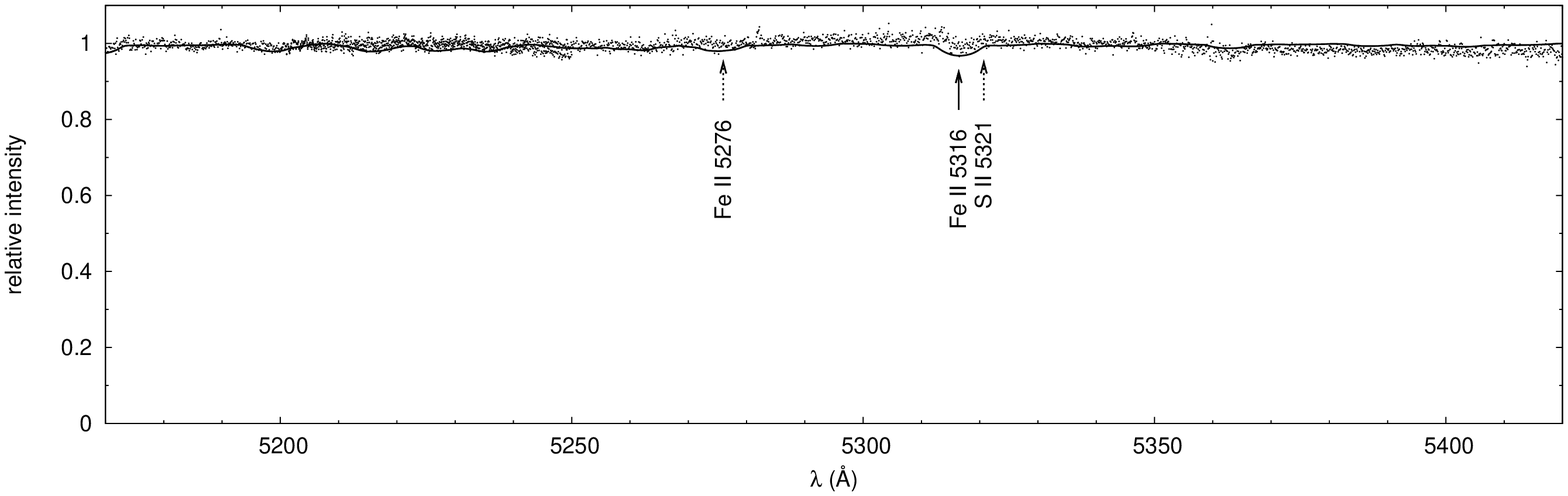}
\includegraphics[width=\hsize]{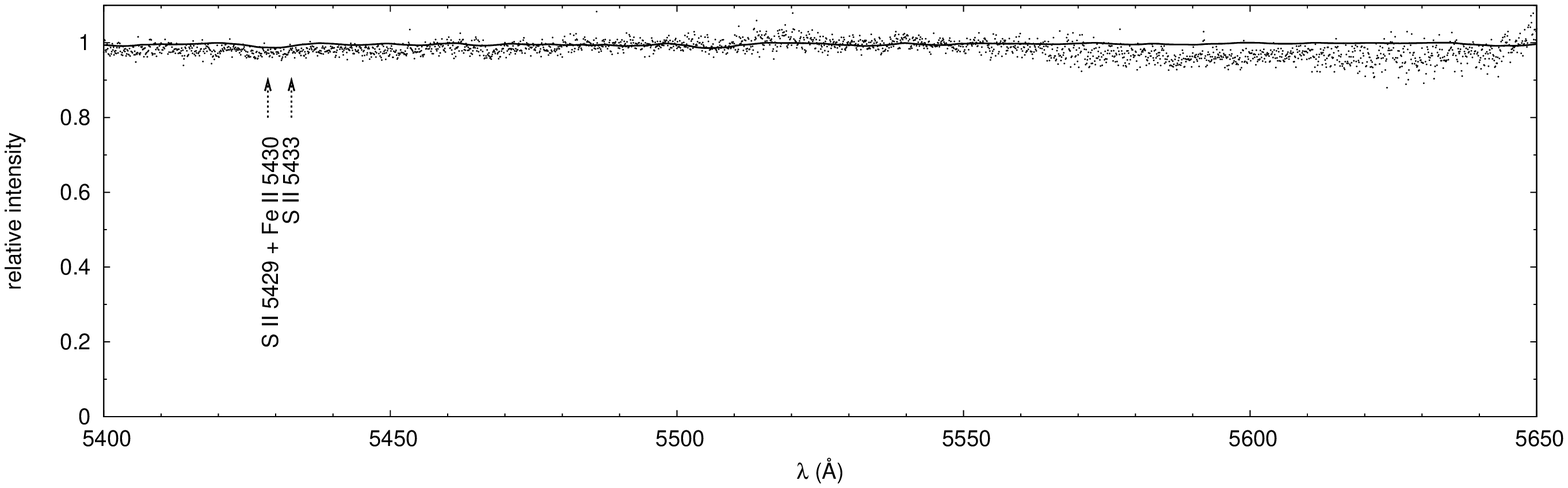}
\caption{contd.}
\end{figure*}

\begin{figure*}
\includegraphics[width=\hsize]{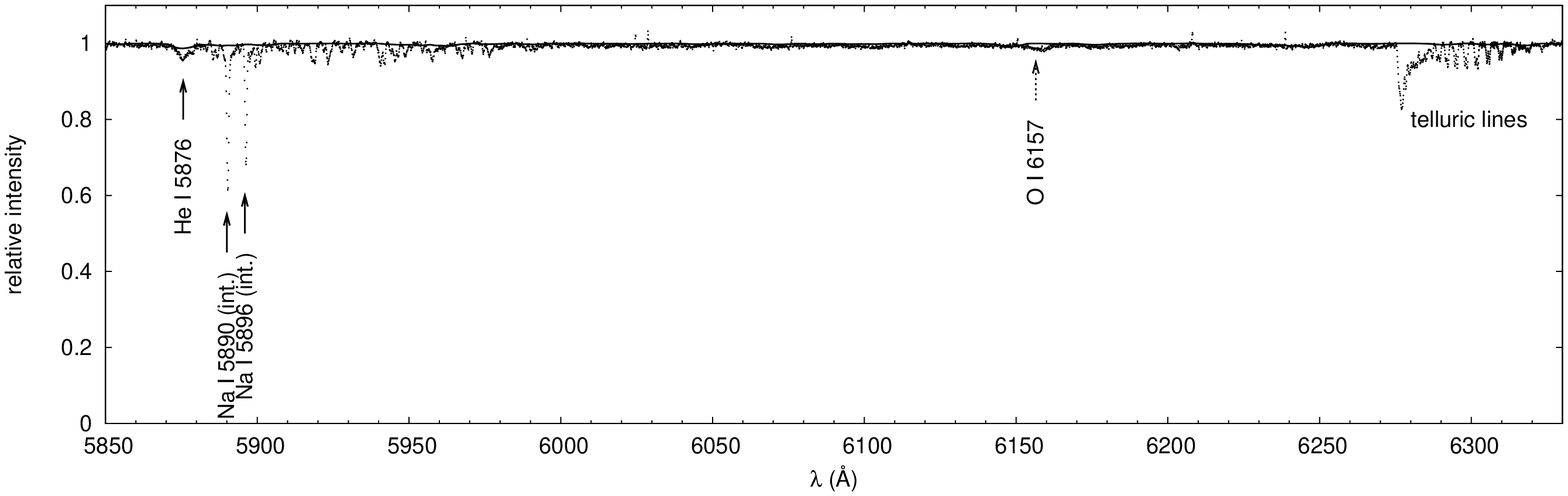}
\includegraphics[width=\hsize]{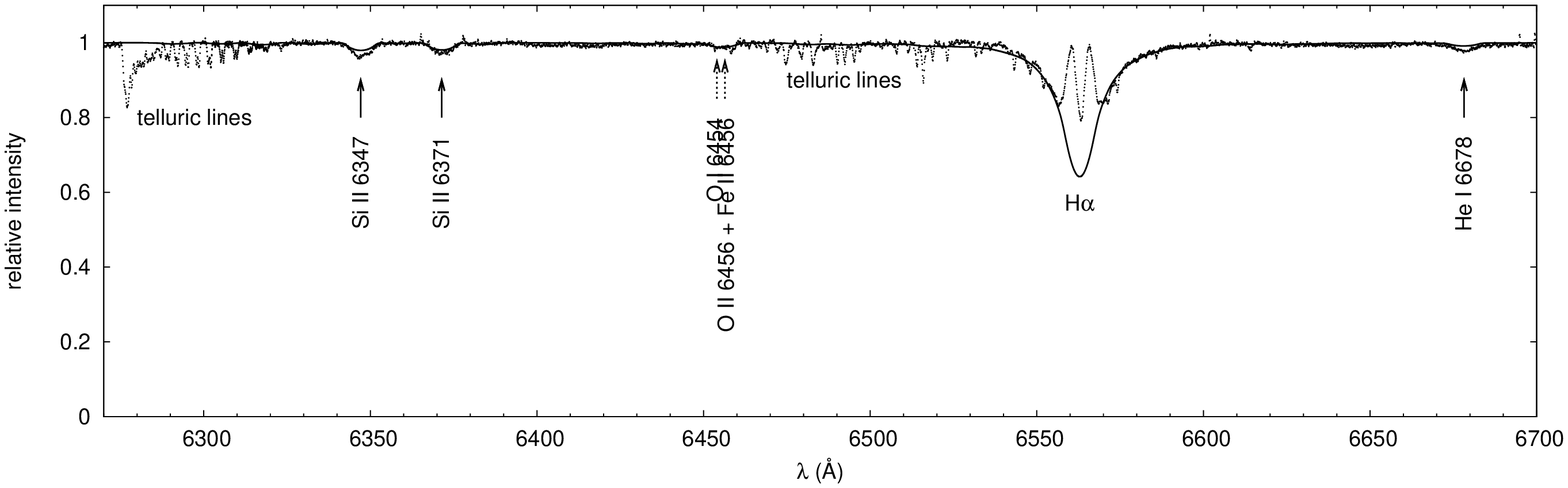}
\includegraphics[width=\hsize]{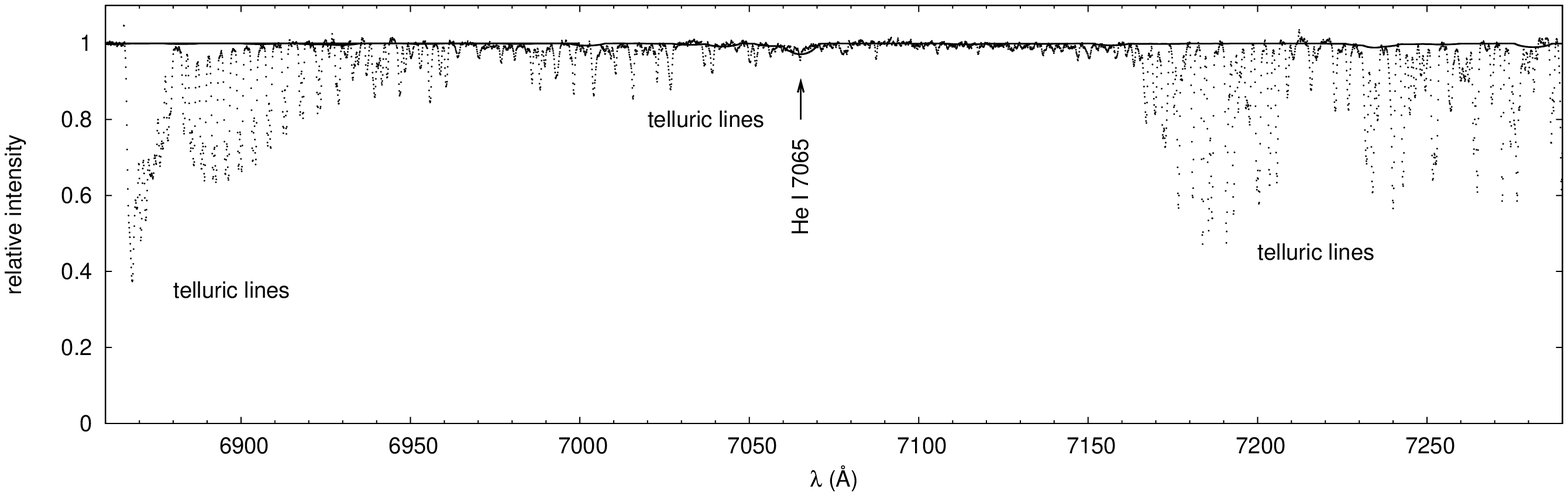}
\includegraphics[width=\hsize]{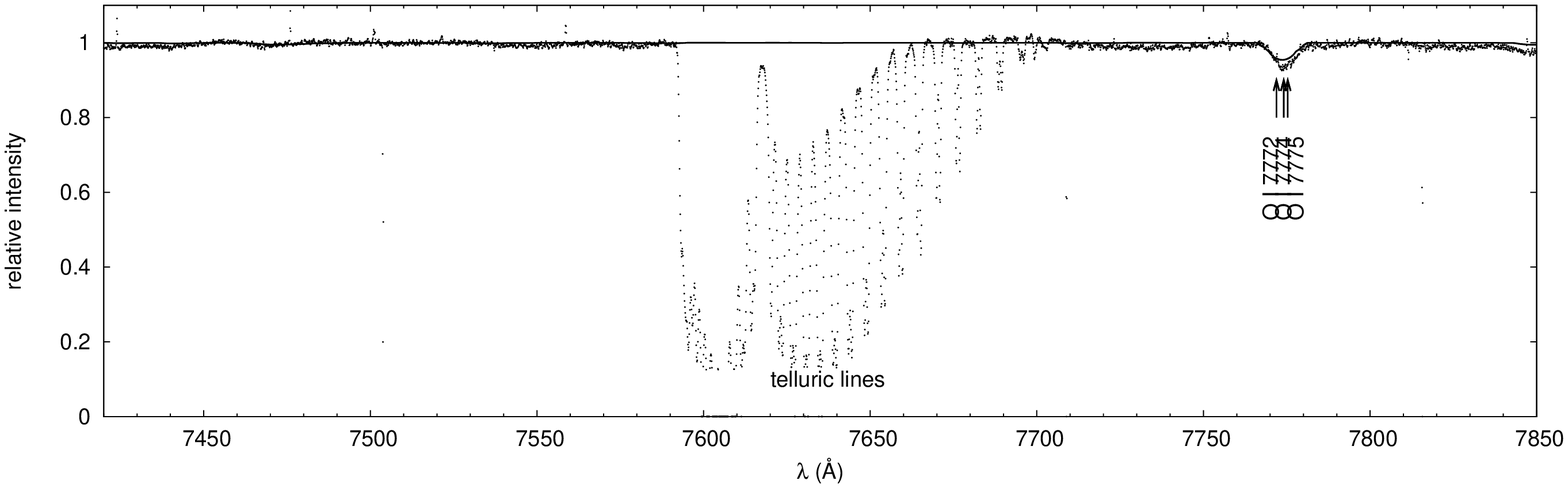}
\caption{Line identification of the spectrum of {\hveb} obtained with
the red channel of the {\HEROS} spectrograph (dots), and comparison with
a synthetic spectrum calculated from the \cite{Kur13} LTE model
atmosphere $\Teff=12\,500\Kelvin$, $\log g=3.5$, rotationally broadened
with $\vsinib = 230\kms$ (full line).
Note the presence of telluric lines and bands near $6280${\AA},
{\Halpha}, $6870${\AA}, $7250${\AA}, and $7590${\AA.}}
\label{HERBred}
\end{figure*}

\begin{table*}[h]
\caption{Heliocentric radial velocities (\RV) of {\hvea} -- spectra from
Ond\v{r}ejov Observatory, where $V_\mathrm{hel}$ is the heliocentric
velocity.}
\label{rvaond}
\begin{center}
\begin{tabular}{llrrrrrrr} \hline
file & HJD & $V_\mathrm{hel}$ &
{\RV} (\Halpha) &
{\RV} (\ion{He}{i} 6678\,\AA) &
{\RV} (\ion{Si}{ii} 6347\,\AA) &
{\RV} (\ion{Si}{ii} 6371\,\AA) \\
& (JD$-$2400000) & (\kms) & (\kms) & (\kms) & (\kms) & (\kms) \\
\hline
\tt ra5788  & 52721.3728&$-$29.71&      &      &      &      \\
\hline
\tt mj170023& 52930.5671&   25.36& 21.45& 21.11& 18.65& 20.07\\	
\tt mj170025& 52930.6315&   25.23& 21.03& 20.75& 17.80& 18.98\\
\tt mk120038& 52956.5608&   16.02& 21.45& 21.67& 17.67& 20.76\\
\tt ml090033& 52983.4198&    3.03& 20.12& 18.98& 16.15& 18.86\\
\tt na050019& 53010.4552&$-$11.13& 13.55& 15.33& 14.84& 16.80\\
\tt na050021& 53010.5053&$-$11.23& 16.48& 15.82& 13.41& 16.43\\  
\tt nd210018& 53117.2973&$-$23.56& 13.44& 16.89& 15.73& 16.70\\ 
\tt nd210019& 53117.3210&$-$23.57& 15.78& 17.09& 13.39& 12.61\\  	
\tt nd220014& 53118.3330&$-$23.35& 15.06& 18.36& 13.79& 16.32\\	
\tt nd300002& 53126.3175&$-$20.47& 12.67& 13.76& 14.90& 18.87\\
\tt ng300048& 53217.5797&   21.00& 18.70& 10.85& 15.09& 18.95\\	
\tt ng300049& 53217.5864&   21.01& 16.35& 12.35& 12.79& 17.14\\	
\tt nh220044& 53240.5571&   27.37& 15.60& 13.35& 14.20& 13.74\\
\tt nh220045& 53240.5649&   27.36& 14.43& 13.58& 13.35& 15.67\\
\tt nh220049& 53240.5862&   27.35& 15.00& 14.95& 12.49& 14.69\\	
\tt nh220053& 53240.6052&   26.32& 14.99& 10.90& 13.32& 15.52\\	
\tt nh230063& 53241.5625&   27.55& 12.98& 14.95& 13.63& 17.10\\
\tt ni020057& 53251.6021&   28.97& 15.21& 15.56& 11.75& 15.84\\
\tt ni060034& 53255.6179&   28.14& 11.90& 11.14&  9.90& 14.66\\
\tt ni080043& 53257.5953&   29.47& 14.39&  --- & 14.39&  --- \\
\tt nj240062& 53303.5864&   23.00& 11.76& 10.54&  9.57& 12.05\\  	
\tt nk240049& 53334.5637&   10.01&  8.79&  --- &  ---  & --- \\ 	
\tt nl100020& 53350.2773&    2.36& 13.55& 11.30&  9.52& 11.50\\	
\tt oa160018& 53387.3315&$-$16.40& 12.44& 12.31& 10.18& 11.89\\
\tt oc220010& 53452.3157&$-$29.61& 10.35& 10.01&  6.21&  5.04\\
\tt oc310022& 53461.3659&$-$28.62& 10.47& 10.03&  6.91& 10.42\\
\tt od010015& 53462.2998&$-$28.42& 11.33&  9.78&  6.25& 11.64\\
\tt od010016& 53462.3100&$-$28.41& 10.15& 10.12&  5.51& 10.18\\ 	
\tt oi060036& 53620.5833&   29.34&  9.90&  9.15&  7.15&  4.46\\ 	
\tt oi060037& 53620.5954&   29.33&  9.30& 11.79&  8.23& 10.97\\  	
\tt oj070018& 53651.4465&   27.84& 23.89& 21.65& 20.52& 20.50\\ 	
\tt oj070028& 53651.4953&   27.79& 23.28& 21.63& 21.83& 23.74\\ 	
\tt oj070034& 53651.5630&   27.69& 24.35& 21.76& 20.64& 22.79\\ 	
\tt oj070040& 53651.6182&   27.58& 23.69& 21.45& 19.95& 21.99\\
\tt oj070042& 53651.6418&   27.54& 23.05& 23.36& 20.76& 21.94\\ 	
\tt oj070044& 53651.6703&   27.48& 25.93& 24.46& 21.79& 23.70\\ 	
\tt oj080058& 53652.4732&   27.62& 25.34& 26.92& 25.08& 25.40\\    	
\tt oj080062& 53652.5220&   27.56& 23.91& 23.42& 21.44& 23.02\\  	
\tt oj080069& 53652.5815&   27.46& 21.98& 23.37& 20.87& 22.44\\  	
\tt oj080073& 53652.6229&   27.38& 24.64& 28.68& 25.40& 23.28\\  	
\tt oj080079& 53652.6823&   27.26& 27.72& 25.32& 23.06& 27.61\\ 	
\tt oj260033& 53670.4313&   21.60& 26.45& 24.66& 19.67& 23.76\\	
\tt oj260035& 53670.4447&   21.58& 24.58& 22.92& 21.83& 22.17\\	
\tt oj290036& 53673.4859&   20.51& 25.52& 22.75& 20.12& 21.90\\	
\tt oj290037& 53673.5018&   20.48& 21.98& 22.49& 20.33& 22.87\\	
\tt pa080032& 53744.3096&$-$12.56& 16.11& 16.95& 14.12& 14.79\\ 	
\tt pj110024& 54020.6019&   26.67& 11.60& 11.71&  9.29& 12.49\\
\tt qc160019& 54176.2664&$-$29.87&  9.03& 10.00&  5.16&  8.53\\
\tt qc160037& 54176.3253&$-$29.98&  8.83& 10.50&  6.73&  7.65\\
\tt qc160046& 54176.4145&$-$30.04&  9.94&  9.52&  6.06&  8.80\\
\tt qc270011& 54187.3995&$-$29.25&  8.95& 10.44&  5.00&  7.82\\
\tt qd020010& 54193.3620&$-$28.37&  7.03& 10.44&  4.54&  7.95\\
\tt qd070002& 54198.3054&$-$27.40&  9.50& 10.71&  6.31&  8.86\\ 
\tt qd130019& 54204.2939&$-$26.03&  8.63&  9.98&  3.03&  7.47\\
\tt qd130023& 54204.3337&$-$26.05&  9.20&  9.15&  5.43&  4.80\\
\tt qd140022& 54205.3262&$-$25.79&  9.31&  9.81&  5.81&  8.30\\
\tt qe020011& 54223.3037&$-$20.03& 13.92&  --- &  --- &  --- \\
\tt qe020012& 54223.3097&$-$20.03& 11.02& 12.46&  2.89&  6.97\\
\tt qh240029& 54337.5997&   27.59& 11.53& 13.34& 12.39& 15.14\\
\tt qh250033& 54338.6177&   27.71& 11.84& 11.91&  9.87& 13.95\\
\tt qi150036& 54359.5365&   29.70& 20.28& 22.21& 18.66& 20.35\\
\tt qi160031& 54360.5360&   29.70& 20.50& 22.06& 19.61& 23.03\\
\tt qi160039& 54360.6371&   29.57& 21.48& 20.63& 17.17& 21.79\\
\tt qj130036& 54387.4787&   26.51& 21.26& 18.68& 21.17& 20.95\\
\tt qj130040& 54387.5534&   26.39& 22.08& 20.12& 20.21& 21.70\\
\tt qj130046& 54387.6826&   26.13& 23.03& 21.13& 19.13& 21.96\\
\tt rb250029& 54522.3202&$-$28.60& 17.53& 16.29& 15.33& 17.04\\
\hline
\end{tabular}
\end{center}
\end{table*}

\begin{table}[h]
\caption{Heliocentric radial velocities (\RV) of {\hvea} -- spectra from
Rozhen Observatory.}
\label{rvaroz}
\begin{center}
\begin{tabular}{llrr}
\hline
\multicolumn{1}{l}{file} & HJD &
\multicolumn{1}{r}{$V_\mathrm{hel}$} & {\RV} (\Halpha) \\ 
& (JD$-$2400000) & \multicolumn{1}{r}{(\kms)} & (\kms) \\
\hline
\tt 04j119& 53303.4455&   23.17 & 10.52 \\
\tt 04j120& 53303.4563&   23.16 &  9.14 \\
\tt 04j303& 53306.5530&   21.99 & 11.16 \\
\tt 04j304& 53306.5638&   21.97 & 12.06 \\
\tt 04k109& 53333.3218&   10.99 &  8.39 \\
\tt 05c062& 53452.6889&$-$29.64 &  4.76 \\
\tt 05c163& 53453.3861&$-$29.56 &  7.58 \\
\tt 05c235& 53454.4102&$-$29.47 &  4.93 \\
\tt 07h050& 54341.5910&   28.25 & 11.92 \\
\tt 07h129& 54342.5379&   28.43 & 12.12 \\
\tt 07h135& 54342.6011&   28.36 & 12.05 \\
\tt 07h215& 54343.5844&   28.52 & 12.19 \\ 
\hline
\end{tabular}
\end{center}
\end{table}

\begin{table*}[h]
\caption{Heliocentric radial velocities (\RV) of {\hveb} -- spectra from
Ond\v{r}ejov Observatory.}
\label{rvbond}
\begin{center}
 \begin{tabular}{llrrrrrrr} \hline
file & HJD & $V_\mathrm{hel}$ & \multicolumn{3}{c}{{\RV} (\Halpha)} &
\\
& (JD$-$2400000) & (\kms) & (\kms) & (\kms) & (\kms) \\
& & & wings & emission & absorption \\
\hline
\tt ra5806  & 52721.8197&$-$29.59& 11.10 &  9.73 & 14.30 \\
\hline
\tt mj170025& 52930.6315&   25.23& 15.40 & 12.47 & 18.92 \\
\tt mk120038& 52956.5608&   16.02& 15.05 & 12.71 & 19.70 \\
\tt nc290015& 53094.3765&$-$28.89& 14.39 & 13.21 & 17.90 \\
\tt nd300001& 53126.3043&$-$20.50& 18.17 & 16.41 & 18.17 \\
\tt nh220047& 53240.5755&   27.33& 27.44 & 16.00 & 12.75 \\
\tt ni020058& 53251.6217&   27.36&  7.47 &  6.88 &  9.22 \\
\tt ni130036& 53262.5828&   29.65& 10.24 &  9.06 & 10.82 \\
\tt nj240060& 53303.5565&   23.02&  3.83 & -0.19 &  2.16 \\
\tt nj240061& 53303.5692&   22.99&  8.58 &  7.40 &  7.99 \\
\tt oa160016& 53387.3113&$-$16.39& 19.80 &  8.08 & 18.92 \\
\tt od010017& 53462.3213&$-$28.44& 10.92 &  9.16 & 10.92 \\
\tt oi060038& 53620.6195&   29.30& 21.00 & 18.06 & 24.51 \\
\tt oi220034& 53636.6211&   29.34& 23.59 & 24.17 & 24.76 \\
\tt oi220035& 53636.6365&   29.31& 17.11 & 20.04 & 23.52 \\
\tt oj070034& 53651.5630&   27.69& 19.70 & 17.94 & 29.08 \\
\tt oj070040& 53651.6182&   27.58& 12.55 &$-$2.11 & 13.14 \\
\tt oj070042& 53651.6418&   27.54& 21.86 &  30.69 & 21.89 \\
\tt oj070044& 53651.6703&   27.48& 21.25 &  14.80 & 27.70 \\
\tt oj080060& 53652.4977&   27.60&  ---  &  37.52 & 24.62 \\
\tt oj280042& 53672.5077&   21.87& 12.90 &  16.42 & 19.94 \\
\tt oj290035& 53673.4623&   21.60& 11.50 &   7.39 &  ---  \\
\tt pj110026& 54020.6245&   26.62&  7.44 &  19.17 & 13.31 \\
\tt qc160019& 54176.2664&$-$29.87&  8.43 &   7.26 & 10.78 \\
\tt qc160039& 54176.3427&$-$29.98&  7.94 &   5.60 & 11.46 \\
\tt qc160048& 54176.4353&$-$30.05&  8.46 &   7.87 & 11.39 \\
\tt qc270010& 54187.3763&$-$29.24&  2.51 &   7.78 & 11.30 \\
\tt qd020011& 54193.3770&$-$28.37&  5.14 &  22.14 &  3.97 \\
\tt qd130021& 54204.3139&$-$26.05&  8.04 &   3.93 & 11.56 \\
\tt qd140024& 54205.3461&$-$25.80&  4.30 &   8.99 &  7.23 \\
\tt qh240030& 54337.6245&   27.57& 20.89 &  32.61 & 23.23 \\
\tt qi150038& 54359.5627&   29.68& 22.60 &  22.02 & 21.43 \\
\tt qi160033& 54360.5562&   29.68& 23.68 &  37.17 & 24.27 \\
\tt qi160039& 54360.6371&   29.57& 20.12 &  37.13 & 20.12 \\
\tt qj130046& 54387.6826&   26.13& 22.66 &  32.04 & 23.25 \\
\tt rb250027& 54522.2805&$-$28.54&  8.89 \\
\tt rb250033& 54522.3822&$-$28.73&  8.21 \\
\tt rb250035& 54522.4158&$-$28.77&  7.35 \\
\tt rc220012& 54548.3622&$-$29.65& 21.58 \\
\hline
\end{tabular} 
\end{center}
\end{table*}

\begin{table*}[h]
\caption{Heliocentric radial velocities (\RV) of {\hveb} -- spectra from
Rozhen Observatory.}
\label{rvbroz}
\begin{center}
\begin{tabular}{llrrrrrrr} \hline
file & HJD & $V_\mathrm{hel}$ & \multicolumn{3}{c}{ {\RV} (\Halpha)} &
\\
& (JD$-$2400000) & (\kms) & (\kms) & (\kms) & (\kms) \\
& & & wings & emission & absorption \\
\hline
\tt 07h049  & 54341.5732&   28.25& 21.98 &  35.69 & 23.52 \\
\tt 07h136  & 54342.6176&   28.33& 21.19 &  28.47 & 26.19 \\
\tt 07h216  & 54343.5993&   28.50& 21.33 &  25.90 & 21.33 \\
\hline
\end{tabular} 
\end{center}
\end{table*}

\end{document}